\documentclass{article}

\usepackage{arXiv}

\usepackage[utf8]{inputenc} 
\usepackage[T1]{fontenc}    
\usepackage{hyperref}       
\usepackage{url}            
\usepackage{booktabs}       
\usepackage{amsfonts}       
\usepackage{nicefrac}       
\usepackage{microtype}      
\usepackage{lipsum}		
\usepackage{graphicx}
\usepackage{doi}

\usepackage[numbers,sort&compress]{natbib}
\title{A thin plate approximation for ocean wave interactions with an ice shelf}

\author{Luke G.~Bennetts \\
	University of Adelaide, Australia \\
	\texttt{luke.bennetts@adelaide.edu.au} 
	\And
	Timothy~D.~Williams \\ 
	NERSC, Norway \\
	\texttt{timothy.williams@nersc.no}
	\And
	Richard~Porter \\ 
	University of Bristol, UK \\
	\texttt{richard.porter@bristol.ac.uk}
}

\date{}

\renewcommand{\shorttitle}{Thin plate approximation for ice shelves}

\hypersetup{
pdftitle={Thin plate approximation for ice shelves},
pdfsubject={q-bio.NC, q-bio.QM},
pdfauthor={Luke G.~Bennetts, Timothy~D.~Williams, Richard~Porter},
pdfkeywords={First keyword, Second keyword, More},
}

\usepackage{amsmath}

\newcommand{\mx}{\textrm{mx}}

\newcommand{\littleo}{\textrm{o}}

\renewcommand{\d}{\,\mathrm{d}} 	

\newcommand{\wrt}{\,\mathrm{d}}   
    
\newcommand{\ci}{\mathrm{i}}   
\newcommand{\e}{\mathrm{e}}

\usepackage{enumerate}

\usepackage{overpic,rotating}

\newcommand{\tn}[1]{\textnormal{#1}}

\newcommand{\beq}{\begin{equation}}
\newcommand{\eeq}{\end{equation}}
\newcommand{\bea}{\begin{eqnarray}}
\newcommand{\eea}{\end{eqnarray}}
\newcommand{\bean}{\begin{eqnarray*}}
\newcommand{\eean}{\end{eqnarray*}}

\newlength{\mylinelength}
\newlength{\mydashlength}
\newlength{\mydashspace}
\newlength{\mychainlengthA}
\newlength{\mychainlengthB}
\newlength{\mychainspace}
\newlength{\mylinethickness}
\newcommand{\mathand}{\quad\textnormal{and}\quad}
\newcommand{\mathandR}{\textnormal{and}\quad}  
  
\newcommand{\mathfor}{\quad\textnormal{for}\quad}

\newcommand{\mathwhere}{\quad\textnormal{where}\quad}
\newcommand{\mathwhereR}{\textnormal{where}\quad}

\newcommand{\mathXL}[1]{\quad\textnormal{#1}}

\usepackage{pifont}
\ifdefined\showmylabels
 \usepackage[inline]{showlabels}
 
\fi

\usepackage[normalem]{ulem}
\usepackage{soul}
\usepackage{tcolorbox}
\ifdefined\shownewold
   \newcommand{\del}[1]{{\color{red}\sout{#1}}}
\else
   \newcommand{\del}[1]{\ignorespaces}
\fi
\ifdefined\shownewold
   \newcommand{\deleqn}[1]{\\\del{\parbox{\textwidth}{#1}}}
\else
   \newcommand{\deleqn}[1]{\ignorespaces}
\fi
\ifdefined\shownewold
   \newcommand{\new}[1]{{\color[rgb]{0,0,1}#1}}
\else
   \newcommand{\new}[1]{#1}
\fi
\ifdefined\shownewold
   \newcommand{\old}[1]{{\color[rgb]{0.5,0.5,0.5}#1}}
\else
   \newcommand{\old}[1]{\ignorespaces}
\fi
\ifdefined\shownewold
   \newcommand{\delref}[1]{\del{\parbox{\figwidth\columnwidth}{#1}}}
\else
   \newcommand{\delref}[1]{\vspace{-12pt}}
\fi
\ifdefined\shownewold
   \newcommand{\lu}[1]{{\color[rgb]{1,0,1} #1}}
   \newcommand{\luchange}[2]{\del{#1}\new{#2}}
   \newcommand{\luleft}[1]{{\color[rgb]{1,0,1}$\longleftarrow$#1}}
   \newcommand{\luright}[1]{{\color[rgb]{1,0,1}#1$\longrightarrow$}}
\else
   \newcommand{\lu}[1]{\ignorespaces}
   \newcommand{\luchange}[1]{\ignorespaces}
   \newcommand{\luleft}[1]{\ignorespaces}
   \newcommand{\luright}[1]{\ignorespaces}
\fi
\usepackage{tikz,pgfplots}
\usetikzlibrary{calc,patterns,decorations.pathmorphing,decorations.markings,decorations.pathmorphing,angles,quotes,arrows,arrows.meta}
\tikzset{snake it/.style={decorate, decoration=snake}}

\newlength{\figurewidth}
\newlength{\figwidth}
\newlength{\figureheight}

\usepackage{xcolor}
\definecolor{darkgreen}{rgb}{0,0.55,0}
\definecolor{midgreen}{rgb}{0,0.8,0.2}
\definecolor{magenta}{rgb}{1,0,1}
\definecolor{purple}{rgb}{0.5,0,0.5}
\definecolor{darkorange}{rgb}{1,0.55,0}
\definecolor{maroon}{rgb}{0.5,0,0}
\definecolor{olive}{rgb}{0.5,0.5,0}
\definecolor{midgrey}{rgb}{0.5,0.5,0.5}
\definecolor{lightgrey}{rgb}{0.75,0.75,0.75}
\definecolor{matlabblue}{rgb}{0,0.447,0.741}
\definecolor{matlabred}{rgb}{0.85,0.325,0.098}
\definecolor{lightblue}{rgb}{0,0.5,1}
\definecolor{darkgrey}{rgb}{0.25,0.25,0.25}
\definecolor{teal}{rgb}{0,0.5,0.5}
\definecolor{navy}{rgb}{0,0,0.5}
\definecolor{goldenrod}{rgb}{0.85,0.6,0.1}

\begin{document}
\maketitle

\begin{abstract}
	A variational principle is proposed to derive the governing equations for the problem of ocean wave interactions with a floating ice shelf, where the ice shelf is modelled by the full linear equations of elasticity and has an Archimedean draught.
	The variational principle is used to form a thin-plate approximation for the ice shelf, which includes water--ice coupling at the shelf front and extensional waves in the shelf, in contrast to the benchmark thin-plate approximation for ocean wave interactions with an ice shelf.
	The thin-plate approximation is combined with a single-mode approximation in the water, where the vertical motion is constrained to the eigenfunction that supports propagating waves.
	The new terms in the approximation are shown to have a major impact on predictions of ice shelf strains for wave periods in the swell regime. 
\end{abstract}


%
\section{Introduction} \label{sec:intro}

For over half a century, thin floating elastic plates have been the benchmark model for ocean wave-induced flexural motions of sea ice \cite{evans1968wave,wadhams1988attenuation,meylan1994response,vaughan_decay_2009,montiel2016attenuation,pitt2022model}
	and ice shelves \cite{holdsworth1978iceberg,vinogradov1985oscillation,fox_squire91,williams2007wave,papathanasiou2015hydroelastic,meylan2021swell}. 
The benchmark model, which dates back to \citet{greenhill1916skating}, assumes the vertical ice displacements are uniform with respect to thickness (i.e., a thin plate), and the water is a potential-flow fluid.
The plate appears in the model through flexural and inertial restoring forces at the water surface, which are manifested as high-order derivatives in the dynamic surface condition. 
This results in so-called flexural-gravity waves, as well as wave modes that have no analogue in open water (i.e., where the water surface is in contact with air), which are typically oscillatory-decaying waves but can become purely decaying in certain regimes \cite{williams2006reflections,bennetts2007wave}.

In both sea ice and ice shelf applications, the canonical wave--ice interaction problem involves a two-dimensional water domain (one horizontal dimension plus depth), which has half of its surface covered by ice, and where motions are excited by an incident wave from the open (non-ice covered) water \cite{evans1968wave,tkacheva2001scattering,linton2003reflection}.
The incident wave is partially reflected at the ice edge and partially transmitted into the ice-covered domain.
The model is used to predict, e.g., strains in landfast sea ice \cite{fox_squire91,fox1994oblique} and ice shelves \cite{fox1991coupling}, and is the basis for models of wave attenuation in the marginal ice zone \cite{bennetts2012calculation,bennetts2012model}.
Although the Archimedean draught of ice is $\approx{}90$\% of its thickness, the thinness of sea ice  been used to justify the so-called shallow-draught approximation, in which the ice floats at the water surface with no submergence.
Therefore, the ice edge experiences no loading, and free edge conditions are applied (i.e., zero bending moment and shear stress).
The water and ice are coupled along the underside of the ice only.

Methods have been developed to accommodate Archimedean ice draught, whilst retaining the free edge conditions \cite{williams2009effect,montiel2012transient,papathanasiou2019resonant}.
The methods address the geometrical corner created by the partial submergence of the ice edge, but not the additional water--ice coupling  created by the bending moment applied by the water motion on the ice edge and the kinematic coupling between the ice edge and the water.
Notably, \citet{porter2004approximations} (and subsequently \citet{bennetts_multi-mode_2007} who corrected an error in \cite{porter2004approximations}) derived the incorrect free edge conditions as the natural conditions of a variational principle, but where the thinness of the plate was already applied in the underlying Lagrangian, i.e., a one dimensional body was partially submerged in a two-dimensional fluid. 

Although ad-hoc, the use of the shallow-draught approximation and/or free edge conditions at a sea ice edge intuitively seems unlikely to have a major impact on the model predictions, as the ice thickness (typically tens of centimetres to a few metres) is much smaller than other characteristic lengths.
Relevant wavelengths are in the swell regime (tens to hundreds of metres; wave periods 10--30\,s) and wave--sea ice interactions typically  occur in the deep ocean ($>1$\,km, i.e., much greater than wavelengths). 
In contrast, ice shelves are hundreds of metres thick, occur on continental shelves and the sub-ice shelf water cavities are typically hundreds of metres deep. 
Ice shelves vibrate in response to ocean waves from long swell (wavelengths order hundreds of metres) to infragravity waves (kilometres to tens of kilometres; wave periods 50--300\,s) and longer \cite{chen2019ross}.
Therefore, the jump in water depth created by the ice draught affects model predictions \cite{kalyanaraman2019shallow}.

For the ice shelf application, water--ice coupling  at the submerged portion of the shelf front (i.e., the ice edge) appears likely to influence model predictions for incident swell.
Compelling evidence that swell forced shelf front strains strong enough to trigger runaway ice shelf disintegrations \cite{massom_etal18} makes this missing aspect of the benchmark thin-plate model particularly conspicuous. 
\citet{abrahams2023ocean} recently analysed a numerical time domain simulation, in which the ice shelf is modelled using the full (linear) equations of elasticity.
In addition to flexural waves, they identified extensional waves in the shelf that are generated by water--ice coupling at the shelf front.
Further, they showed that extensional wave displacement amplitudes exceed those of the flexural waves for low frequencies, with the extensional to flexural amplitude ratio tending to infinity as the frequency tends to zero.
\citet{kalyanaraman_coupled_2020} analysed numerical computations in the frequency domain of an ice shelf (of finite length) modelled using the full equations of elasticity (although neglecting gravity), but applied free edge conditions at the shelf front.
They found that the flexural displacement profiles were similar to those predicted by the benchmark model, at least for two wave periods in the infragravity regime.  
The finding is broadly consistent with the results of studies using the shallow-draught approximation and thick plate models \cite{fox1991coupling,balmforth1999ocean}.

In this article, we outline a variational principle that derives the governing equations of the ice shelf problem, where the shelf has an  an Archimedean draught and is modelled by the full equations of elasticity.
We use the variational principle to derive a thin-plate equation through a process of averaging with respect to the shelf thickness, which includes extensional waves in the shelf and water--ice coupling at the shelf front.
We combine the thin-plate approximation with a single-mode approximation in the water, which involves averaging with respect to depth, similar to \cite{porter2004approximations,bennetts_multi-mode_2007}.
We use the approximations to investigate the influence of coupling at the ice edge and extensional waves on ice shelf strains, across the swell and infragravity wave regimes.


\section{Preliminaries}\label{sec:prelims}

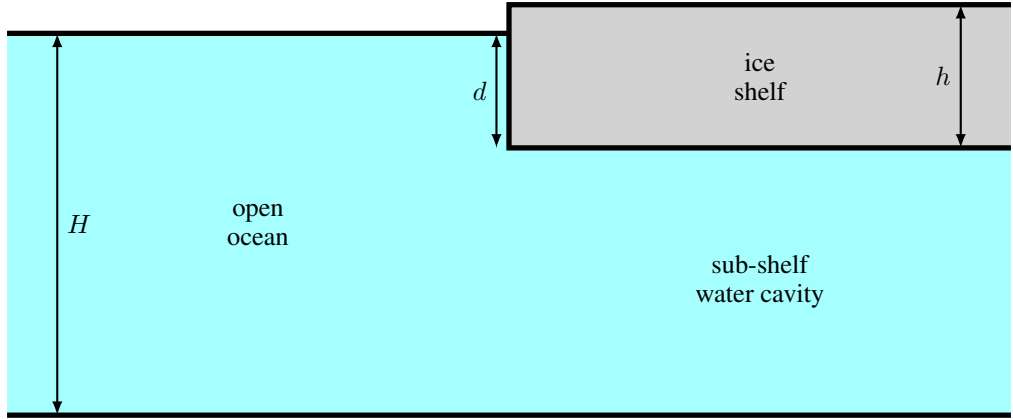
\begin{figure}[h!]
        \centering
        \setlength{\figurewidth}{0.85\textwidth} 
        \setlength{\figureheight}{0.4\textwidth} 
        \def\labelsA{1}
        \def\labelsB{1}
%
\definecolor{mycolor1}{rgb}{0.00000,1.00000,1.00000}%

\def\L{3000} 
\def\H{200}  
\def\d{60}   
\def\f{15}   

\begin{tikzpicture}

\begin{axis}[%
width=0.951\figurewidth,
height=\figureheight,
at={(0\figurewidth,0\figureheight)},
scale only axis,
xmin=-\L,
xmax=\L,
xtick={},
xlabel style={font=\color{white!15!black}},
xticklabels={},
xlabel={$x$},
ymin=-210,
ymax=50,
ytick={-200,0},
yticklabels={$-h_{0}$,0},
ylabel style={font=\color{white!15!black}},
ylabel={$z$},
axis background/.style={fill=white},
hide axis
]

\coordinate (origin) at (axis cs:0,0);
\coordinate (e1) at (axis cs:1,0);
\coordinate (e2) at (axis cs:0,1);

\coordinate (A) at (axis cs:-\L,-\H);
\coordinate (B) at (axis cs:0,-\H);
\coordinate (C) at (axis cs:\L,-\H);
\coordinate (D) at (axis cs:\L,-\d);
\coordinate (E) at (axis cs:0,-\d);
\coordinate (F) at (axis cs:-\L,0);
\coordinate (G) at (axis cs:0,\f);
\coordinate (H) at (axis cs:\L,\f);
\coordinate (I) at (axis cs:-\L,-\d);

\fill[mycolor1!35, domain=0:4000, variable=\x, samples=200] (A) -- (B) -- (C) -- (D) -- (E) -- (origin) -- (F) --cycle;

\fill[gray!35, line width=2.0pt, domain=0:4000, variable=\x, samples=200] (D) -- (E) -- (G) -- (H) -- cycle;

\draw[black, line width=2.0pt] (A) -- (B) -- (C); 
\draw[black, line width=2.0pt] (D) -- (E) -- (G) -- (H);  
\draw[black, line width=2.0pt, domain=0:4000, variable=\x, samples=200] (F) -- (origin);

\if\labelsA1
\node[align=center,xshift=0] at ($(E)!0.5!(H)$) {ice \\ shelf};
\node[align=center,xshift=0] at ($(C)!0.5!(E)$) {sub-shelf \\ water cavity}; 
\node[align=center] at ( $ (A)!0.5!(origin) $ ) {open \\ ocean}; 
\fi
\if\labelsB1
\draw[latex-latex,black,thick] ($(B)!0.9!(A)$) -- node[right] {$H$} ($(origin)!0.9!(F)$);
\draw[latex-latex,black,thick] ($(E)!0.025!(I)$) -- node[left] {$d$} ($(origin)!0.025!(F)$);
\draw[latex-latex,black,thick] ($(E)!0.9!(D)$) -- node[left] {$h$} ($(G)!0.9!(H)$);
\fi



\end{axis}


\end{tikzpicture}%
        \caption{Schematic (not to scale) of the undeformed geometry.}
        \label{fig:schematic-finite-uniform}
\end{figure}

Consider a two-dimensional domain of inviscid and irrotational water, which has an (undisturbed) finite depth $H$ and infinite horizontal extent (Fig.~\ref{fig:schematic-finite-uniform}).
An ice shelf of finite thickness $h$ and semi-infinite length covers the surface of the right-hand side of the water domain.
Let the Cartesian coordinate system $(x,z)\equiv{}(x_{1},x_{2})$ define locations in the water and ice shelf. 
The horizontal coordinate, $x\in\mathbb{R}$, has its origin set to coincide with the shelf front.
The vertical coordinate, $z$, has its origin set to coincide with the undisturbed water surface, such that the (flat) bed is located at $z=-H$.
The ice shelf is assumed to have an Archimedean draught, such that its (undisturbed) lower surface is located at 
\begin{equation}
z=-d\equiv\frac{\rho_{\tn{i}}\,h}{\rho_{\tn{w}}},	
\end{equation}
where $\rho_{\tn{i}}=922.5$\,kg\,m$^{-3}$ and $\rho_{\tn{w}}=1025$\,kg\,m$^{-3}$ are the ice and water densities, respectively, such that $\rho_{\tn{i}}\,/\,\rho_{\tn{w}}=0.9$.
The ice/water domain is partitioned into the ice shelf, the sub-shelf water cavity and the open ocean (Fig.~\ref{fig:schematic-finite-uniform}), respectively,
\begin{subequations}
\begin{align}
\Omega_{\tn{sh}} & =  \{(x,z): 0<x<\infty; -d<z<h-d\}
\\[10pt]
\Omega_{\tn{ca}} & = \{(x,z): 0<x<\infty; -H<z<-d\},
\\[10pt]
\mathandR
\Omega_{\tn{op}} & = \{(x,z): -\infty<x<0; -H<z<0\}.
\end{align}	
\end{subequations}

Small amplitude (linear) motions of the ice--water system are considered.
Let the displacement field be 
\beq
\mathbf{u}(x,z,t)=\left[U(x,z,t);W(x,z,t)\right]\equiv[U_{1}(x,z,t);U_{2}(x,z,t)].
\eeq
In the ice, the stress and strain tensors are $\boldsymbol{\sigma}(x,z,t)$ and $\boldsymbol{\varepsilon}(x,z,t)$, respectively, with components
\begin{subequations}\label{eqn:stress_strain}
\beq	
\sigma_{ij} = \lambda\,\delta_{ij}\sum_{r=1}^{2}\varepsilon_{rr} + 2\,\mu\,\varepsilon_{ij}
	\mathand
	\varepsilon_{ij} = \frac{1}{2}\left(U_{j,x_{i}} + U_{i,x_{j}}\right)
	\mathfor
	i,j=1,2,
	\tag{\theequation a,b}
\eeq
\end{subequations}
where $\lambda$ and $\mu$ are the Lam\'{e} coefficients.
In the water, the stress tensor has components
\beq
\sigma_{ij}=-P\,\delta_{ij},
\eeq
where $P(x,z,t)$ is the pressure field, 
and the displacement field is expressed in terms of a displacement potential $\Phi(x,z,t)$, such that
\beq
\mathbf{u}=\nabla\Phi.
\eeq
The vertical displacements of the free surfaces  are $\zeta_{\bullet}(x,t)$ for $\bullet=\tn{w--a},\tn{w--i},\tn{i--a}$, which correspond to the water--atmosphere, water--ice and ice--atmosphere interfaces, respectively.


The relative hydrostatic pressures in the open ocean, ice shelf, sub-shelf water cavity and open ocean are (respectively)
\begin{subequations}\label{eqs:pressures}
\begin{equation}
    P_{\tn{op}}(z) = - \rho_{\tn{w}}\,g\,z,
    \quad
    P_{\tn{sh}}(z) = P_{0} - \rho_{\tn{i}}\,g\,(z+d),
    \mathand
    P_{\tn{ca}}(z) = P_{0} - \rho_{\tn{w}}\,g\,(z+d) = -\rho_{\tn{w}}\,g\,z,
    \tag{\theequation a,b,c}
\end{equation}
\end{subequations}
where $P_0=P_{\tn{sh}}(-d)=P_{\tn{ca}}(-d)=P_{\tn{op}}(-d)= \rho_{\tn{i}}\,g\,h=\rho_{\tn{w}}\,g\,d$, and $g=9.81$\,m\,s$^{-2}$ is the constant gravitational acceleration. 
Note that $P_{\tn{sh}}(h-d)=P_{\tn{op}}(0)=0$, so Eq.~(\ref{eqs:pressures}a--c) represent the true hydrostatic pressure minus the constant atmospheric pressure, $P_{\tn{at}}$, and that the hydrostatic pressure is continuous going from the open ocean into the sub-shelf cavity.

\section{Variational principle}\label{sec:vp}

\subsection{Lagrangian}

The Lagrangian for the ice--water system is
\beq
\mathcal{L}\{\mathbf{u},\Phi,P,\boldsymbol{\zeta},\boldsymbol{\tau}\}
=
\mathcal{L}_{\tn{sh}}\{\mathbf{u},\zeta_{\text{i--a}},\zeta_{\text{w--i}},\boldsymbol{\tau}\}
+
\mathcal{L}_{\tn{ca}}\{\Phi,P,\zeta_{\text{w--i}},\boldsymbol{\tau}\}
+
\mathcal{L}_{\tn{op}}\{\Phi,P,\zeta_{\text{w--a}},\boldsymbol{\tau}\},
\eeq
where $\mathcal{L}_{\tn{sh}}$, $\mathcal{L}_{\tn{ca}}$ and $\mathcal{L}_{\tn{op}}$ are the Lagrangians for the ice shelf, sub-shelf water cavity and open ocean, respectively.


The (linearised)  Lagrangian for the ice shelf is expressed as $\mathcal{L}_{\tn{sh}}=\mathcal{T}_{\tn{sh}}-\mathcal{V}_{\tn{sh}}$, where $\mathcal{T}_{\tn{sh}}$ and $\mathcal{V}_{\tn{sh}}$ are the kinetic and potential energies in the ice shelf, respectively.
The kinetic energy is
\begin{subequations}\label{eqs:Lshelf}
\begin{align}
\mathcal{T}_{\tn{sh}} \{\mathbf{u}\}
& = 
\frac{\rho_{\tn{i}}}{2}
\int_{t_{0}}^{t_{1}} \iint_{\Omega_{\tn{sh}}} \left\{ U_{t}^{2} + W_{t}^{2}\right\} \wrt{}x\wrt{z}\wrt{}t,
\end{align}
\new{where $t_{0}$ and $t_{1}$ are arbitrary times, such that $t_{1}>t_{0}$.}
The potential energy is the integral of the strain energy density plus the gravitational potential over the shelf domain, plus integrals from linearisation of the moving boundaries and stresses applied to the boundaries (denoted $\tau_{ij}$), so that
\begin{align}\nonumber
\mathcal{V}_{\tn{sh}} \{\mathbf{u},\zeta_{\text{i--a}},\zeta_{\text{w--i}},\boldsymbol{\tau}\}
& =
\int_{t_{0}}^{t_{1}} \iint_{\Omega_{\tn{sh}}} 
\left\{ \frac{1}{2} \sum_{i=1}^{2}\sum_{j=1}^{2}\sigma_{ij}\,\varepsilon_{ij} + P_{\tn{sh}}(z-W) \right\} \wrt{}x\wrt{z}\wrt{}t
\\[8pt] \nonumber 
& ~~ 
+
\int_{t_{0}}^{t_{1}} 
\int_{0}^{\infty} 
\Big[\rho_{\tn{i}}\,g\,W\,\zeta_{\text{i--a}}
- 
\frac12 \, \rho_{\tn{i}}\,g\,\zeta_{\text{i--a}}^2
-
\tau_{22}\,W 
\Big]_{z=h-d}  
\wrt{}x
\wrt{}t
\\[8pt] \nonumber 
& ~~ 
-
\int_{t_{0}}^{t_{1}} 
\int_{0}^{\infty} 
\Big[
(P_{0}+\rho_{\tn{i}}\,g\,W)\,\zeta_{\text{w--i}}
- 
\frac12 \rho_{\tn{i}}\,g\,\zeta_{\text{w--i}}^2
-
\tau_{22}\,W 
\Big]_{z=-d}  
\wrt{}x
\wrt{}t
\\[8pt] \nonumber
& ~~
+
\int_{t_{0}}^{t_{1}} 
\int_{-d}^{h-d} 
\Big[
\tau_{11}\,U
\Big]_{x=0}  
\wrt{}z
\wrt{}t
\\[8pt] 
& ~~
-
\int_{t_{0}}^{t_{1}} 
\int_{-d}^{h-d} 
\Big[ 
\tau_{11}\,U 
\Big]_{x\to\infty}  
\wrt{}z
\wrt{}t
.
\end{align}
\end{subequations}


The Lagrangian for the sub-shelf water cavity is expressed as
$\mathcal{L}_{\tn{ca}}=\mathcal{T}_{\tn{ca}}-\mathcal{V}_{\tn{ca}}$, where 
the kinetic energy in the water cavity is
\begin{subequations}\label{eqs:Lcavity}
\begin{align} 
\mathcal{T}_{\tn{ca}} \{\Phi\}
& =  
\frac{\rho_{\tn{w}}}{2}
\int_{t_{0}}^{t_{1}} \iint_{\Omega_{\tn{ca}}} \left\{ \Phi_{xt}^{2} + \Phi_{zt}^{2}\right\} \wrt{}x\wrt{z}\wrt{}t.
\end{align}
The strain energy density in the water is defined in terms of the pressure and potential, so that the
potential energy is
\begin{align}\nonumber
\mathcal{V}_{\tn{ca}} \{\Phi,P,\zeta_{\text{w--i}},\boldsymbol{\tau}\}
& =
\int_{t_{0}}^{t_{1}} \iint_{\Omega_{\tn{ca}}} 
\left\{ 
-P\,\nabla^{2}\Phi 
+ 
P_{\tn{ca}}(z-\Phi_z)
\right\} \wrt{}x\wrt{z}\wrt{}t
\\[8pt] \nonumber 
& ~~ +
\int_{t_{0}}^{t_{1}} 
\int_{0}^{\infty} 
\Big[(P_0 + \rho_{\tn{w}}\,g\,\Phi_z)\,\zeta_{\text{w--i}}
- 
\frac12\, \rho_{\tn{w}}\,g\,\zeta_{\text{w--i}}^2
-
\tau_{22}\,\Phi_z 
\Big]_{z=-d}  
\wrt{}x
\wrt{}t
\\[8pt] \nonumber 
& ~~ 
-
\int_{t_{0}}^{t_{1}} 
\int_{0}^{\infty} 
\Big[
-\tau_{22}\,\Phi_z 
\Big]_{z=-H}  
\wrt{}x
\wrt{}t
\\[8pt] 
& ~~
-
\int_{t_{0}}^{t_{1}} 
\int_{-H}^{-d} 
\Big[\tau_{11}\,\Phi_x 
\Big]_{x=0}^{\infty}  
\wrt{}z
\wrt{}t.
\end{align}
\end{subequations}


Similarly, the linearised Lagrangian for the open ocean is 
$\mathcal{L}_{\tn{op}}=\mathcal{T}_{\tn{op}}-\mathcal{V}_{\tn{op}}$, in which
\begin{subequations}\label{eqs:Lopen}
\begin{align} 
\mathcal{T}_{\tn{op}} \{\Phi\}
& =  
\frac{\rho_{\tn{w}}}{2}
\int_{t_{0}}^{t_{1}} \iint_{\Omega_{\tn{op}}} \left\{ \Phi_{xt}^{2} + \Phi_{zt}^{2}\right\} \wrt{}x\wrt{z}\wrt{}t
\end{align}
\begin{align}\nonumber
\mathandR 
\mathcal{V}_{\tn{op}} \{\Phi,P,\zeta_{\text{w--a}},\boldsymbol{\tau}\}
& =
\int_{t_{0}}^{t_{1}} \iint_{\Omega_{\tn{op}}} 
\left\{ 
-P\,\nabla^{2}\Phi 
+ 
P_{\tn{op}}(z-\Phi_z)
\right\} \wrt{}x\wrt{z}\wrt{}t
\\[8pt] \nonumber 
& ~~ 
+
\int_{t_{0}}^{t_{1}} 
\int_{-\infty}^{0} 
\Big[\rho_{\tn{w}}\,g\,\Phi_z\,\zeta_{\text{w--a}}
- 
\frac12\, \rho_{\tn{w}}\,g\,\zeta_{\text{w--a}}^2
-
\tau_{22}\,\Phi_z 
\Big]_{z=0}  
\wrt{}x
\wrt{}t
\\[8pt] \nonumber 
& ~~
-
\int_{t_{0}}^{t_{1}} 
\int_{-\infty}^{0} 
\Big[
-\tau_{22}\,\Phi_z 
\Big]_{z=-H}  
\wrt{}x
\wrt{}t
\\[8pt] 
& ~~ 
+
\int_{t_{0}}^{t_{1}} 
\int_{-H}^{0} 
\Big[
-\tau_{11}\,\Phi_x 
\Big]_{x\to-\infty}^{0} 
\wrt{}z
\wrt{}t.
\end{align}
\end{subequations}


Small variations are applied to all unknowns, such that the Lagrangians become
\begin{equation}
	\mathcal{T}_{\tn{sh}}\{\mathbf{u}+\delta\mathbf{u}\}
	=
	\mathcal{T}_{\tn{sh}}\{\mathbf{u}\}
	+
	\delta \mathcal{T}_{\tn{sh}}\{\mathbf{u}:\delta{}\mathbf{u}\}
	+
	\littleo(\delta{}\mathbf{u}),
	\mathXL{and so on.}
\end{equation}
The first variation of the full Lagrangian, $\delta\mathcal{L}\{\mathbf{u},\Phi,P,\boldsymbol{\zeta},\boldsymbol{\tau}:\delta\mathbf{u},\delta\Phi,\delta{}P,\delta\boldsymbol{\zeta},\delta\boldsymbol{\tau}\}$, is
\begin{equation}
\delta\mathcal{L} =
\delta\mathcal{L}_{\tn{sh}}+\delta\mathcal{L}_{\tn{ca}}+\delta\mathcal{L}_{\tn{op}}	
=
\delta\mathcal{T}_{\tn{sh}}-\delta\mathcal{V}_{\tn{sh}}
+\delta\mathcal{T}_{\tn{ca}}-\delta\mathcal{V}_{\tn{ca}}
+\delta\mathcal{T}_{\tn{op}}	-\delta\mathcal{V}_{\tn{op}}.
\end{equation}
From Eqs.~(\ref{eqs:Lshelf}--\ref{eqs:Lopen}), the first variation is evaluated as 
\begin{align}\nonumber
\delta\mathcal{L}
& =  
-\int_{t_{0}}^{t_{1}} \iint_{\Omega_{\tn{sh}}} 
\Big\{ 
\delta{}U\,(\rho_{\tn{i}}\,U_{tt} - \sigma_{11,x} - \sigma_{12,z})  
+ 
\delta{}W\,(\rho_{\tn{i}}\,W_{tt} - \sigma_{21,x} - \sigma_{22,z} + \rho_{\tn{i}}\,g)
\Big\} \wrt{}x\wrt{z}\wrt{}t
\\[8pt] \nonumber
&~~
+
\int_{t_{0}}^{t_{1}} \iint_{\Omega_{\tn{ca}}} 
\left\{ 
\delta{}P\,\nabla^{2}\Phi 
+ 
\delta{}\Phi\,\nabla^{2}\hat{P} 
\right\} 
\wrt{}x\wrt{z}\wrt{}t
\\[8pt] \nonumber
& ~~
+
\int_{t_{0}}^{t_{1}} \iint_{\Omega_{\tn{op}}} 
\left\{ 
\delta{}P\,\nabla^{2}\Phi 
+ 
\delta{}\Phi\,\nabla^{2}\hat{P} 
\right\} 
\wrt{}x\wrt{z}\wrt{}t
\\[8pt] \nonumber 
&  ~~ -
\int_{t_{0}}^{t_{1}} 
\int_{0}^{\infty} 
\Big[ \delta\zeta_{\text{i--a}}\,\rho_{\tn{i}}\,g\,(W-\zeta_{\text{i--a}})
+
\delta{}W\,(\sigma_{22}+\rho_{\tn{i}}\,g\,\zeta_{\text{i--a}}
	+P_{\tn{at}}) 
+ 
\delta{}U\,\sigma_{12} 
\Big]_{z=h-d}  
\wrt{}x
\wrt{}t
\\[8pt] \nonumber
& ~~
+
\int_{t_{0}}^{t_{1}} 
\int_{0}^{h-d} 
\Big[
\delta{}W\,\sigma_{12} 
+
\delta{}U\,(\sigma_{11}+P_{\tn{at}}) 
\Big]_{x=0}  
\wrt{}z
\wrt{}t
\\[8pt] \nonumber 
&  ~~ +
\int_{t_{0}}^{t_{1}} 
\int_{0}^{\infty} 
\Big[
\delta\zeta_{\text{w--i}}\,\{\rho_{\tn{i}}\,g\,(W-\zeta_{\text{w--i}})-\rho_{\tn{w}}\,g\,(\Phi_{z}-\zeta_{\text{w--i}})\}
+
\delta{}W\,(\sigma_{22}+\rho_{\tn{i}}\,g\,\zeta_{\text{w--i}}-S_{\tn{bt}}) 
+ 
\delta{}U\,\sigma_{12}
\\[6pt] \nonumber
& ~~~~~~~~~~~~~~~
-
\delta\Phi\,\hat{P}_{z}
+
\delta\Phi_{z}\,(P-\rho_{\tn{w}}\,g\,\zeta_{\text{w--i}}+S_{\tn{bt}}) 
-
\delta{}S_{\tn{bt}}\,(W-\Phi_{z}) 
\Big]_{z=-d}  
\wrt{}x
\wrt{}t
\\[8pt] \nonumber
& ~~
+
\int_{t_{0}}^{t_{1}} 
\int_{-d}^{0} 
\Big[
\delta{}W\,\sigma_{12} 
+
\delta{}U\,(\sigma_{11}-S_{\tn{fr}}) 
-
\delta\Phi\,\hat{P}_{x}
+
\delta\Phi_{x}\,(P+S_{\tn{fr}}) 
-
\delta{}S_{\tn{fr}}\,(U-\Phi_{x})
\Big]_{x=0}  
\wrt{}z
\wrt{}t
\\[8pt]  \nonumber
& ~~ 
+
\int_{t_{0}}^{t_{1}} 
\int_{-\infty}^{\infty} 
\Big[ 
\delta\Phi\,\hat{P}_{z}
-
\delta\Phi_{z}\,(P+S_{\tn{bd}}) 
-
\delta{}S_{\tn{bd}}\,\Phi_{z}
\Big]_{z=-H}  
\wrt{}x
\wrt{}t
\\[8pt] \nonumber 
& ~~ 
-
\int_{t_{0}}^{t_{1}} 
\int_{-\infty}^{0} 
\Big[ 
\delta\zeta_{\text{w--a}}\,\rho_{\tn{w}}\,g\,(\Phi_{z}-\zeta_{\text{w--a}})
+
\delta\Phi\,\hat{P}_{z}
-
\delta\Phi_{z}\,(P-\rho_{\tn{w}}\,g\,\zeta_{\text{w--a}}
-P_{\tn{at}}) 
\Big]_{z=0}  
\wrt{}x
\wrt{}t
\\[8pt] 
\label{eqn:deltaL}
& ~~
+
\int_{t_{0}}^{t_{1}} 
\int_{-H}^{-d} 
\Big[ 
\delta\Phi\,\hat{P}_{x}
-
\delta{}\Phi_{x} \, \big(P+S_{\tn{fr}}\big) 
\Big]_{x=0^-}^{0^+} 
\wrt{}z
\wrt{}t
-
\int_{t_{0}}^{t_{1}} 
\int_{-H}^{-d} 
\Big[ 
\delta{}S_{\tn{fr}}\,\langle\Phi_{x}\rangle
\Big]_{x=0} 
\wrt{}z
\wrt{}t
.
\end{align}
Here, $\langle\bullet\rangle$ denotes the jump in the included quantity over $x=0$, and the notations
\begin{subequations}
\begin{align}
\hat{P}(x,z,t) &\equiv P + \rho_{\tn{w}}\,\left(\Phi_{tt} + g\,z\right),
&\quad
S_{\tn{fr}}(z,t) & \equiv
[\tau_{11}]_{x=0},
\tag{\theequation a,b}
\\[8pt]
S_{\tn{bt}}(x,t)
&\equiv
[\tau_{22}]_{z=-d}
&\mathand
S_{\tn{bd}}(x,t)
&\equiv
[\tau_{22}]_{z=-H},
\tag{\theequation c,d}
\end{align}
\end{subequations}
have been introduced for convenience\new{, where the subscripts fr, bt and bd indicate stresses on the shelf front, shelf bottom and seabed, respectively.}.
The known atmospheric stresses have been incorporated, such that
\beq
[\tau_{22}]_{x<0,z=0}
=
[\tau_{11}]_{x=0,0<z<h-d}
=
[\tau_{22}]_{x>0,z=h-d}
\equiv
-P_{\tn{at}},
\eeq
so that their first variations vanish.
All variations are assumed to vanish in the far field $x\to\pm\infty$.


\subsection{Governing equations}


Enforcing $\delta{}\mathcal{L}=0$ for arbitrary variations, $\delta\mathbf{u}$ and so on,
$\hat{P}$ must satisfy Laplace's equation
\begin{equation}\label{eq:Phat_laplace}
	\nabla^{2}\hat{P} = 0 
	\mathfor 
	(x,z)\in\Omega_{\tn{op}}
	\mathand
	(x,z)\in\Omega_{\tn{ca}},
\end{equation}
(from domain integral terms proportional to $\delta\Phi$ in Eq.~\ref{eqn:deltaL}), 
with boundary conditions
\begin{subequations}\label{eqs:Phat_bcs}
\begin{align}
	\hat{P}_{x} &=0
	\mathfor
	x=0,\;-d<z<0,
	&\quad 
	\hat{P}_{z}&=0
	\mathfor
	-\infty<x<0,\;z=0,
	\tag{\theequation a,b}
	\\[8pt]
	\hat{P}_{z}&=0
	\mathfor
	0<x<\infty,\;z=-d,
	&\mathand
	\hat{P}_{z}&=0
	\mathfor
	-\infty<x<\infty,\;z=-H,
	\tag{\theequation c,d}
\end{align}
\end{subequations}
(from the terms proportional to $\delta\Phi$ in the respective boundary integrals).

Eqs.~(\ref{eq:Phat_laplace}--\ref{eqs:Phat_bcs}) for $\hat{P}$ are uncoupled from the other unknowns, and can be solved to give 
\begin{subequations}\label{eqs:Phat}
\begin{equation}
	\hat{P}=C_{\tn{op}}(t) \mathfor (x,z)\in\Omega_{\tn{op}}
	\mathand
	\hat{P}=C_{\tn{ca}}(t) \mathfor (x,z)\in\Omega_{\tn{ca}}, 
	\tag{\theequation a,b}
\end{equation}
\end{subequations}
where $C_{\tn{op}}$ and $C_{\tn{ca}}$ are arbitrary functions.

Water pressures in $\Omega_{\tn{op}}$ and $\Omega_{\tn{ca}}$ can be deduced from Eqs.~(\ref{eqs:Phat}a,b), respectively. 
If we also set 
\begin{equation}
C_{\tn{op}}=C_{\tn{ca}}\equiv{}P_{\tn{at}},
\end{equation}
(implicitly using the freedom in the potential $\Phi$), the water pressure is given by Bernoulli's equation as the sum of the hydrostatic pressure (introduced earlier) and a dynamic pressure, such that
\begin{equation}\label{eq:bernoulli_pressure}
    P=P_{\tn{at}}-\rho_{\tn{w}}\,\left(\Phi_{tt} + g\,z\right)
	\mathfor
	(x,z)\in\Omega_{\tn{op}}\cup\Omega_{\tn{ca}}.
\end{equation}



From the remaining conditions given by $\delta{}\mathcal{L}=0$, it is possible to deduce the \textit{field equations of the full linear problem:}
\begin{tcolorbox}[width=\linewidth, colback=white!95!yellow] 
\begin{subequations}\label{eqs:full_field}
\begin{align}
\rho_{\tn{i}}\,U_{tt} & =\sigma_{11,x} + \sigma_{12,z}
&& \mathfor (x,z)\in  \Omega_{\tn{sh}},
\\[8pt]
\rho_{\tn{i}}\,W_{tt} & = \sigma_{12,x} + \sigma_{22,z}-\rho_{\tn{i}}\,g
&& \mathfor (x,z)\in  \Omega_{\tn{sh}},
\\[8pt]
\mathand
\nabla^{2}\Phi & = 0
&& \mathfor (x,z)\in \Omega_{\tn{ca}}\cup\Omega_{\tn{op}},
\end{align}
\end{subequations}
\end{tcolorbox}
where the continuities at the ocean--cavity interface $\langle\Phi\rangle=\langle\Phi_{x}\rangle=0$ have been used.
\new{Eqs.~(\ref{eqs:full_field}a,b) are the full equations of linear elasticity in the ice shelf, and Eq.~(\ref{eqs:full_field}c) is Laplace's equation potential in the water, resulting from the standard assumptions of potential flow theory. }

Further, $\delta{}L=0$ derives the \textit{interfacial equations of the full linear problem:}
\begin{tcolorbox}[width=\linewidth, colback=white!95!yellow] 
\begin{subequations}\label{eqs:BCs_full}
\begin{align}
W=\zeta_{\text{i--a}},
\quad
\sigma_{12}=0
\mathand
\sigma_{22}+\rho_{\tn{i}}\,g\,\zeta_{\text{i--a}} = -P_{\tn{at}} 
&
\mathfor
0<x<\infty
,\;
z=h-d,
\tag{\theequation a,b,c}
\\[8pt]
\sigma_{12}=0
\mathand
\sigma_{11}= - P_{\tn{at}} 
&
\mathfor
x=0
,\;
0<z<h-d,
\tag{\theequation d,e}	
\\[8pt]\nonumber
	W=\Phi_{z}=\zeta_{\text{w--i}},
	\quad
	\sigma_{12}=0,
	\quad
	\sigma_{22}+\rho_{\tn{i}}\,g\,\zeta_{\text{w--i}} = S_{\tn{bt}} &
\\[8pt]
	\mathandR
	P-\rho_{\tn{w}}\,g\,\zeta_{\text{w--i}}=-S_{\tn{bt}}
	& \mathfor
	0<x<\infty
	,\;
	z=-d,
	\tag{\theequation f,g,h,i}
\\[8pt]
	U=\Phi_{x},
	\quad
	\sigma_{12}=0	
	\mathand
	S_{\tn{fr}} = -P=\sigma_{11}
	&\mathfor
	x=0 
	,\;
	-d<z<0,
	\tag{\theequation j,k,l}
\\[8pt]
	\Phi_{z}=0
	\mathand
	S_{\tn{bd}}=-P
	&\mathfor
	-\infty<x<\infty
	,\;
	z=-H,
	\tag{\theequation m,n}
\\[8pt]
	\Phi_{z}=\zeta_{\text{w--a}}
	\mathand
	P-\rho_{\tn{w}}\,g\,\zeta_{\text{w--a}} =  P_{\tn{at}} 
	&\mathfor
	x<0
	,\;
	z=0.
	\tag{\theequation o,p}
\end{align}
\end{subequations}
\end{tcolorbox}
\new{Eq.~(\ref{eqs:BCs_full}) contains conditions at the interfaces between (a--e)~the ice shelf and the atmosphere, (f--l)~the ice shelf and the water, (m--n)~the water and the seabed, and (o--p)~the water and the atmosphere.
Eqs.~(\ref{eqs:BCs_full}a,f,j,m,o) are kinematic conditions, i.e., matching of displacements at common boundaries. 
Eqs.~(\ref{eqs:BCs_full}b,d,g,k) are continuities of shear stress (only non-zero in the ice shelf), and Eqs.~(\ref{eqs:BCs_full}c,e,h,i,l,n,p) are continuities of normal stresses.
}



\section{Thin plate approximation}\label{sec:thinplate}

A thin plate (depth averaged)  approximation for the ice shelf displacements, $\mathbf{u}=(U,W)$, is derived using the anzatzes
\begin{subequations}\label{eqn:UW_thinplate}
\begin{equation}
U(x,z,t)\approx{}
\overline{U}(x,t)-(z+d-h\,/\,2)\,\overline{W}_{x}(x,t)
\mathand
W(x,z,t)\approx{}
\overline{W}(x,t),
\tag{\theequation a,b}
\end{equation}
\end{subequations}
which include extensional motions, via $\overline{U}$, as well as flexural motion, via $\overline{W}$.
The components of the strain tensor (\ref{eqn:stress_strain}b) reduce to
\begin{subequations}
\begin{equation}
	\varepsilon_{11}=\overline{U}_{x}-(z+d-h\,/\,2)\,\overline{W}_{xx}
	\mathand 
	\varepsilon_{12}=\varepsilon_{21}=\varepsilon_{22}\equiv{}0.
\tag{\theequation a,b}
\end{equation}
\end{subequations}
All other unknowns are approximated indirectly.

Applying (\ref{eqn:UW_thinplate}) in the ice shelf Lagrangian, $\mathcal{L}_{\tn{sh}}$, its first variation becomes
\begin{align} \nonumber
\delta\mathcal{L}_{\tn{sh}} = & ~ 
-h\,\int_{t_{0}}^{t_{1}}\int_{0}^{\infty}\Big\{\delta{\overline{U}}\, \Big(\rho _{i}\,\overline{U}_{t\,t} - (\lambda+2\,\mu)\, \overline{U}_{x\,x}\Big)\Big\}~\textrm{d}x~\textrm{d}t
\\[10pt] \nonumber & ~
-\int_{t_{0}}^{t_{1}}\int_{0}^{\infty}\Big\{\delta{\overline{W}}\,\Big( \rho_{i}\,h\,\overline{W}_{t\,t}+\frac{h^3\,\{(\lambda+2\,\mu)\,\overline{W}_{x\,x\,x\,x} - \rho_{i}\,\overline{W}_{x\,x\,t\,t}\}}{12} 
\\[6pt] \nonumber
& \qquad\qquad + g\,h\,\rho_{i} + S_{\tn{bt}} 
 	+ P_{\tn{at}} 
	+ g\,\rho_{i}\,(\zeta_{\text{i--a}}-\zeta_{\text{w--i}}) \Big)\Big\}~\textrm{d}x~\textrm{d}t
\\[10pt] \nonumber & ~
+ \int_{t_{0}}^{t_{1}}\int_{0}^{\infty}\Big\{g\,\rho _{i}\,\delta{\zeta_{\text{w--i}}}\,\Big( \overline{W}-\zeta_{\text{w--i}}\Big) 
- 
g\,\rho _{i}\,\delta{\zeta_{\text{i--a}}}\,\Big( \overline{W}-\zeta_{\text{i--a}}\Big)\Big\}~\textrm{d}x~\textrm{d}t
\\[10pt] \nonumber & ~
- \int_{t_{0}}^{t_{1}}\int_{0}^{\infty}\Big\{ \delta{}S_{\tn{bt}}\,\overline{W} \Big\}~\textrm{d}x~\textrm{d}t
\\[10pt] \nonumber & ~
+ \int_{t_{0}}^{t_{1}} \Big[ \delta{\overline{U}}\,\Big( h\,(\lambda+2\,\mu)\, \overline{U}_{x} - \int_{-d}^{h-d}S_{\tn{fr}}~\wrt{}z \Big) \Big]_{x=0} ~\textrm{d}t
\\[10pt] \nonumber & ~ 
+ \int_{t_{0}}^{t_{1}}\left[ \delta{}\overline{W}\,\Big(\frac{\rho_{i}\,h^{3}}{12}\,\overline{W}_{x\,t\,t} - \frac{h^{3}\,(\lambda+2\,\mu)\, \overline{W}_{x\,x\,x}}{12} \Big)\right]_{x=0}~\textrm{d}t
\\[10pt] \nonumber & ~ 
+ \int_{t_{0}}^{t_{1}}\left[ \delta{}\overline{W}_{x}\,\Big( 
\frac{h^{3}\,(\lambda+2\,\mu)}{12}\,\overline{W}_{x\,x} + \int_{-d}^{h-d}\left(d-\frac{h}{2}+z\right)\,S_{\tn{fr}}~\wrt{}z \Big)\right]_{x=0}~\textrm{d}t
\\[10pt] \label{eq:deltaLsh_thin}
& ~ 
- 
\int_{t_{0}}^{t_{1}}\int_{-d}^{0}\Big[\delta S_{\tn{fr}} \, \left(\overline{U}-\left(d-\frac{h}{2}+z\right)\,\overline{W}_{x}-\Phi_x\right)\Big]_{x=0} ~\textrm{d}z~\textrm{d}t.
\end{align}
Combining (\ref{eq:deltaLsh_thin}) with the relevant components of $\delta\mathcal{L}_{\tn{ca}}$, the vertical component of the shelf displacement is coupled to the cavity via the conditions
\begin{subequations}\label{eq:W_Phi_couple}
\begin{align}
\label{eq:shelf_cavity_thinplate_1}
& \overline{W}-\zeta_{\text{i--a}}=0,
\quad
[\Phi_{z}]_{z=-d}-\overline{W}=0,
\quad
P-\rho_{\tn{w}}\,g\,\zeta_{\text{w--i}}+S_{\tn{bt}} = 0,
\tag{\theequation a,b,c}
\\[10pt] \label{eq:shelf_cavity_thinplate_2}
&
g\,\rho _{i}\,\Big( h+\overline{W}-\zeta_{\text{w--i}}\Big) - \Big(P_0+\rho_{\tn{w}}\,g\,([\Phi_{z}]_{z=-d}-\zeta_{\text{w--i}})\Big)=0,
\tag{\theequation d}
\\[10pt] \label{eq:shelf_cavity_thinplate_3}
\mathandR &
\rho_{i}\,h\,\overline{W}_{t\,t}+\frac{h^3\,\{(\lambda+2\,\mu)\,\overline{W}_{x\,x\,x\,x} - \rho_{i}\,\overline{W}_{x\,x\,t\,t}\}}{12} + g\,h\,\rho_{i} + S_{\tn{bt}} + P_{\tn{at}}
	+ g\,\rho_{i}\,(\zeta_{\text{i--a}}-\zeta_{\text{w--i}})=0,
\tag{\theequation e}
\end{align}
\end{subequations}
for $x>0$. 
As $P_{0}=\rho_{i}\,g\,h$,  it follows from (\ref{eq:W_Phi_couple}a,b,d) that
\beq\label{eq:shelf_cavity_thinplate_identity}
\overline{W}=\zeta_{\text{w--i}}=\zeta_{\text{i--a}}=\left[\Phi_{z}\right]_{z=-d}.
\eeq
Substituting \eqref{eq:shelf_cavity_thinplate_identity} into 	(\ref{eq:W_Phi_couple}c,e), and using the Bernoulli pressure \eqref{eq:bernoulli_pressure} and Archimedean draft, 
	 results in a thin plate equation for the ice shelf flexure, forced by the water motion (given below in Eq.~\ref{eqs:thinplate}b).
In contrast, the thin plate equation for the extensional motion (from the first integral in Eq.~\ref{eq:deltaLsh_thin}) is not directly coupled to the cavity.


Therefore, the approximate $\delta{}\mathcal{L}_{\tn{sh}}$ (in Eq.~\ref{eq:deltaLsh_thin}) combined with $\delta{}L_{\tn{ca}}$ derives the \textit{thin plate approximation field equations:}
\begin{tcolorbox}[width=\linewidth, colback=white!95!black,] 
\begin{subequations}\label{eqs:thinplate}
\begin{align}
 \rho_{w}\,\Big([\Phi_{tt}]_{z=-d} + g\,\overline{W}  \Big) +\rho_{i}\,h\,\overline{W}_{t\,t}+\frac{h^3\,\{(\lambda+2\,\mu)\,\overline{W}_{x\,x\,x\,x} - \rho_{i}\,\overline{W}_{x\,x\,t\,t}\}}{12} & =0,
\\[10pt]
\mathandR \rho _{i}\,\overline{U}_{t\,t} - (\lambda+2\,\mu)\, \overline{U}_{x\,x} & = 0, 
\end{align}
\end{subequations}
\end{tcolorbox}
for $x>0$.
\new{Eq.~(\ref{eqs:thinplate}a) is similar to the benchmark thin plate equation, i.e., a Kirchoff plate with fluid loading, but also contains rotational inertia, as with a Timoshelnko-Mindlin plate \cite{fox1991coupling,balmforth1999ocean}.
Eq.~(\ref{eqs:thinplate}b) is the standard field equation for extensional waves in a thin plate \cite{lamb1917waves}.}

Coupling Eq.~(\ref{eq:deltaLsh_thin}) with $\delta\mathcal{L}_{\tn{op}}$ derives the \textit{shelf front conditions for the thin-plate approximation:}
\begin{tcolorbox}[width=\linewidth, colback=white!95!black] 
\begin{subequations}\label{eqs:thinplate_gov}
\begin{align}
	\frac{h^{3}\,(\lambda+2\,\mu)}{12}\,\overline{W}_{x\,x} +
	\int_{-d}^{h-d}\left(d-\frac{h}{2}+z\right)\,S_{\tn{fr}}~\wrt{}z = 0,
	& 
\\[10pt]
	(\lambda+2\,\mu)\, \overline{W}_{x\,x\,x} - \rho_{i}\,\overline{W}_{x\,t\,t}  =0, 
	& 
\\[10pt]
h\,(\lambda+2\,\mu)\, \overline{U}_{x} - \int_{-d}^{h-d}S_{\tn{fr}}~\wrt{}z = 0,
& 
\\[10pt]
\mathandR  \Phi_{x}	-\left(\overline{U}-\left(d-\frac{h}{2}+z\right)\,\overline{W}_{x} \right)  =0
	&
	\mathfor 
	-d<z<0,
\end{align}
\end{subequations}
\end{tcolorbox}
for $x=0$, where $S_{\tn{fr}}=-[P]_{x=0}$ for $z\in(-d,0)$ and $S_{\tn{fr}}=-P_{\tn{at}}$ for $z\in(0,h-d)$. 
Eqs.~(\ref{eqs:thinplate_gov}a--d) represent, respectively, continuity of bending moment, shear stress, normal traction and horizontal displacement. 
\del{Note that (\ref{eqs:thinplate_gov}d) is satisfied only weakly, according to the final integral in (\ref{eq:deltaLsh_thin}) --- this depends on the precise ansatz used for the potential $\Phi$ (see the following section and Appendix~A).}
As in the full linear problem, the potential $\Phi$ satisfies Laplace's equation in the water domain, the impermeable seabed condition and the free surface conditions, i.e., Eq.~(\ref{eqs:full_field}c) and Eqs.~(\ref{eqs:BCs_full}m,o,p).

\section{Frequency domain}\label{sec:frequencydomain}

\subsection{Governing equations and single-mode approximation}\label{sec:sma}

Assume all dynamic components are time-harmonic at some prescribed frequency, $\omega\in\mathbb{R}_{+}$, so that the extensional and flexural components of the ice displacements are, respectively,
\begin{equation}\label{eq:shelf_timeharmo}
\overline{U}(x,t)=u(x)\,\e^{-\ci\,\omega\,t}
\mathand
\overline{W}(x,t)=w(x)\,\e^{-\ci\,\omega\,t},
\end{equation}
and the interfacial displacements are 
\begin{equation}\label{eq:freesurf_timeharmo}
\zeta_{\bullet}(x,t)=\eta_{\bullet}(x)\,\e^{-\ci\,\omega\,t}	
\mathfor \bullet=\tn{w--a}, \tn{w--i}, \tn{i--a},
\end{equation}
where $u,w,\eta_{\bullet}\in\mathbb{C}$ and it is implicitly assumed from here on that only the real parts are retained for the time-dependent variables.
For the water, prescribe Bernoulli pressure via
\begin{equation}
P(x,z,t) =	P_{\tn{at}} - \rho_{\tn{w}}\,\{\Phi_{tt}+g\,z\}
\mathfor (x,z)\in\Omega_{\tn{op}}\cup\Omega_{\tn{ca}}
\quad\Rightarrow\quad
\hat{P}=P_{\tn{at}},
\end{equation}
and constrain the vertical dependence of the potential, such that
\begin{subequations}\label{eqs:SMA}
\begin{align}\label{eq:singlemodeapx_openwater}
\Phi(x,z,t)\approx{}\frac{g}{\omega^{2}}\,\varphi(x)\,\frac{\cosh\{k\,(z+H)\}}{\cosh(k\,H)}	\,\e^{-\ci\,\omega\,t}
\mathfor
(x,z)\in\Omega_{\tn{op}},
\\[12pt]\label{eq:singlemodeapx_flexgrav}
\Phi(x,z,t)\approx{}\frac{g}{\omega^{2}}\,\psi(x)\,\frac{\cosh\{\kappa\,(z+H)\}}{\cosh\{\kappa\,(H-d)\}}	\,\e^{-\ci\,\omega\,t}
\mathfor
(x,z)\in\Omega_{\tn{ca}},
\end{align}
\end{subequations}
for wavenumbers $k$, $\kappa\in\mathbb{R}_{+}$ to be defined, i.e.\ a single-mode approximation \citep{porter2004approximations,bennetts_multi-mode_2007}, noting that Eqs.~(\ref{eqs:SMA}a--b) create a jump in the potential over the interface $z\in(-H,-d)$ for $x=0$.
The stresses at the shelf bottom and front are prescribed as 
\begin{subequations}
\begin{align}
	S_{\tn{bt}}(x) & = -[P]_{z=-d} + \rho_{\tn{w}}\,g\,\zeta_{\tn{w-i}}
	&& \mathfor x>0
	\\[8pt]
	\mathandR 
	S_{\tn{fr}}(z) & = P_{\tn{at}} - \rho_{\tn{w}}\,\{[\Phi_{tt}]_{x=0}+g\,z\}
	&& \mathfor -H<z<0.	
\end{align}
\end{subequations}

Applying these constraints to $\delta{}\mathcal{L}$ in Eq.~\eqref{eqn:deltaL}, using $\delta{}\mathcal{L}_{\tn{sh}}$ in Eq.~\eqref{eq:deltaLsh_thin}, gives
\begin{align} \nonumber
\delta{}\mathcal{L} = &
	-h\,\int_{t_{0}}^{t_{1}}\int_{0}^{\infty}
	\e^{-2\,\ci\,\omega\,t}\,\delta{u}\, 
	\Big\{ 
	-\rho _{i}\,\omega^{2}\,u - (\lambda+2\,\mu)\, u'' \Big\}~\textrm{d}x~\textrm{d}t
\\[10pt] \nonumber & ~
	-\int_{t_{0}}^{t_{1}}\int_{0}^{\infty}
	\e^{-2\,\ci\,\omega\,t}\,\delta{w}\,
	\Big\{
	-\rho_{i}\,h\,\omega^{2}\,w + \frac{h^3\,\{(\lambda+2\,\mu)\,w'''' + \rho_{i}\,\omega^{2}\,w''\}}{12} 
	\\[6pt] \nonumber
	& ~~~~~~~~~~~~~~~~~~~~~~~~~~~~~~~~~~~~~~~~~
	+ g\,\rho_{\tn{w}}\,(\eta_{\text{w--i}}-\psi) 
	+ g\,\rho_{i}\,(\eta_{\text{i--a}}-\eta_{\text{w--i}})\Big\}~\textrm{d}x~\textrm{d}t
	\\[10pt] \nonumber
	&
	+ \frac{\rho_{\tn{w}}\,g^{2}}{\omega^{2}}\,\int_{t_{0}}^{t_{1}} \int_{0}^{\infty} \e^{-2\,\ci\,\omega\,t}\,\delta{}\psi\,\Bigg\{
	\int_{-H}^{-d}
		(\psi''+\kappa^{2}\,\psi)\,\frac{\cosh^{2}\{\kappa\,(z+H)\}}{\cosh^{2}\{\kappa\,(H-d)\}}\wrt{}z
	\\[8pt] \nonumber
	& ~~~~~~~~~~~~~~~~~~~~~~~~~~~~~~~~~~~~~~~~~~~~~~~~~~~~~~~~~
	+ \Big\{\frac{\omega^{2}}{g}\,w-\kappa\,\tanh\{\kappa\,(H-d)\}\,\psi \Big\}\Bigg\}\wrt{}x\wrt{}t
\\[10pt] \nonumber
	& +g\,\int_{t_{0}}^{t_{1}} 
	   \int_{0}^{\infty} \e^{-2\,\ci\,\omega\,t}\,\delta{}\eta_{\tn{w-i}}\,
		({\rho_{\tn{i}}}-{\rho_{\tn{w}}})\,(w-\eta_{\tn{w-i}})
		\wrt{}x\wrt{}t
\\[10pt] \nonumber
	& - \rho_{\tn{i}}\,g\,\int_{t_{0}}^{t_{1}} \int_{0}^{\infty} \e^{-2\,\ci\,\omega\,t}\,\delta{}\eta_{\tn{i-a}}\,(w-\eta_{\tn{i-a}})\wrt{}x\wrt{}t
	\\[10pt] \nonumber
	& + \frac{\rho_{\tn{w}}\,g^{2}}{\omega^{2}}\,\int_{t_{0}}^{t_{1}} \int_{-\infty}^{0} \e^{-2\,\ci\,\omega\,t}\,\delta{}\varphi\,\Bigg\{
	\int_{-H}^{0}
		(\varphi''+k^{2}\,\varphi)\,\frac{\cosh^{2}\{k\,(z+H)\}}{\cosh^{2}(k\,H)}\wrt{}z
	\\[8pt] \nonumber
		& ~~~~~~~~~~~~~~~~~~~~~~~~~~~~~~~~~~~~~~~~~~~~~~~~~~~~~~~~~~~~~~~~~~~~~~~~~~
		+ \tanh(k\,H)\,\{\varphi - \eta_{\tn{w-a}}\}\Bigg\}\wrt{}x\wrt{}t
\\[10pt] 
	& - \frac{\rho_{\tn{w}}\,g^{2}}{\omega^{2}}\,
	    \int_{t_{0}}^{t_{1}} \int_{-\infty}^{0} \e^{-2\,\ci\,\omega\,t}\,\delta{}\eta_{\tn{w-a}}\,
	    \Big(
		k\,\tanh(kH)\,\varphi-\frac{\omega^{2}}{g}\,\eta_{\tn{w-i}}\Big)\wrt{}x\wrt{}z
	+ \delta\mathcal{C}_{\tn{op-ca}} + \delta\mathcal{C}_{\tn{op-sh}},
\end{align}
where $\delta\mathcal{C}_{\tn{op-ca}}$ and $\delta\mathcal{C}_{\tn{op-ca}}$ contain contributions on the interfaces between the open water and the shelf front and cavity, respectively. 

\begin{figure}
\centering
        \setlength{\figurewidth}{0.8\textwidth} 
        \setlength{\figureheight}{0.3\textwidth} 
%
\begin{tikzpicture}

\begin{axis}[%
width=\figurewidth,
height=0.999\figureheight,
at={(0\figurewidth,0\figureheight)},
scale only axis,
xmode=log,
xmin=0.01,
xmax=1,
xminorticks=true,
xtick={0.01,0.1,1},
xticklabels={$-2$,$-1$,$0$},
xlabel style={font=\color{white!15!black}},
xlabel={$\displaystyle\log_{10}\left(\frac{\omega}{2\,\pi}\right)$},
ymode=log,
ymin=1e-06,
ymax=0.2,
ytick={1e-06,1e-05,1e-04,1e-03,1e-02,1e-01},
yticklabels={$-6$,$-5$,$-4$,$-3$,$-2$,$-1$},
yminorticks=true,
axis background/.style={fill=white},
legend style={
at={(current axis.north west)},
legend columns=-1,
/tikz/every even column/.append style={column sep=3mm},
anchor=north west,
xshift=5pt,yshift=-5pt,
draw=white!15!black}
]
\addplot [color=black, line width=1.5pt]
  table[row sep=crcr]{%
0.01	0.000113034562746885\\
0.0199	0.000225849259703839\\
0.0298	0.000340499565979089\\
0.0397	0.000457966357359549\\
0.0496	0.000579313931760228\\
0.0595	0.000705721539741559\\
0.0694	0.000838517192545108\\
0.0793	0.000979212552727255\\
0.0891999999999999	0.00112953561887531\\
0.0991	0.00129145443534886\\
0.109	0.00146717994534371\\
0.1189	0.00165913007183035\\
0.1288	0.0018698332412956\\
0.1387	0.00210175471978961\\
0.1486	0.00235705233698114\\
0.1585	0.00263731122521744\\
0.1684	0.00294334955319436\\
0.1783	0.00327518691354127\\
0.1981	0.00401340086748663\\
0.2278	0.0052920148114322\\
0.3763	0.0144344230404104\\
1	0.101936799184506\\
};
\addlegendentry{~$\log_{10}(k)$}

\addplot [color=matlabblue, line width=1.5pt]
  table[row sep=crcr]{%
0.01	0.000128469812004115\\
0.0199	0.000256930692789323\\
0.0397	0.000517036803211284\\
0.0496	0.000643006126832664\\
0.0595	0.000760103153970518\\
0.0694	0.000865434424736935\\
0.0793	0.000958823548617086\\
0.0891999999999999	0.00104168315381113\\
0.0991	0.00111582192153758\\
0.109	0.00118288147411921\\
0.1189	0.00124419228016278\\
0.1288	0.00130079155909504\\
0.1486	0.00140288358827641\\
0.1684	0.00149368141921731\\
0.1882	0.00157602952792924\\
0.2179	0.00168765506022456\\
0.2575	0.00182017535403256\\
0.3169	0.00199465638726261\\
0.4159	0.00224333849469093\\
0.8218	0.0030012481229196\\
1	0.00326501522812656\\
};
\addlegendentry{~$\log_{10}(\kappa)$}

\addplot [color=matlabred, line width=1.5pt]
  table[row sep=crcr]{%
0.01	3.74048793608535e-06\\
1	0.000374048793608535\\
};
\addlegendentry{~$\log_{10}(q)$}

\end{axis}
\end{tikzpicture}%
\caption{Wavenumbers for the open water ($k$), flexural-gravity wave ($\kappa$) and extensional wave in the shelf ($q$) versus frequency for shelf thickness $h=200$\,m and water depth $H=800$\,m.}	
\label{fig:wavenumbers}
\end{figure}
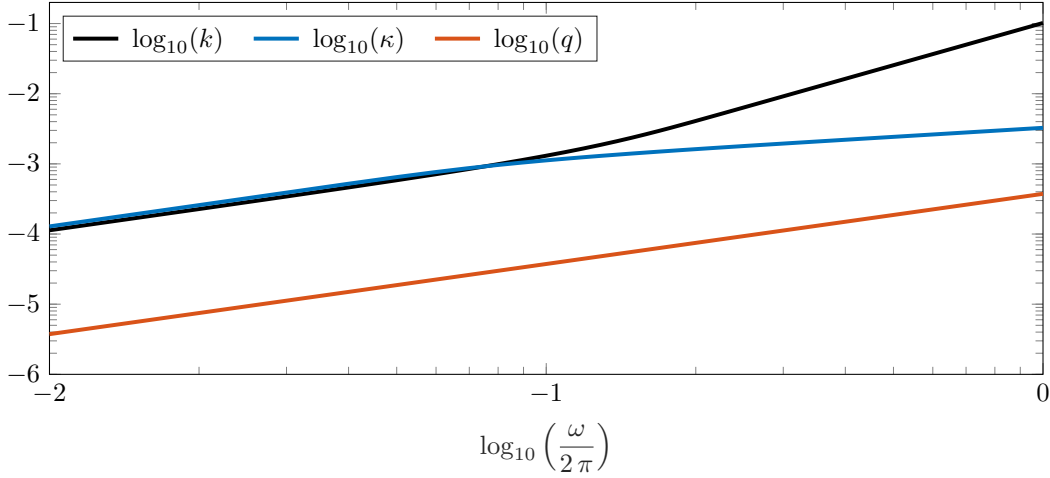

Setting $\delta{}\mathcal{L}=0$ for arbitrary variations ($\delta{}u$ and so on) gives a set of governing equation for the unknown functions of the horizontal spatial coordinate in Eqs.~(\ref{eq:shelf_timeharmo}--\ref{eqs:SMA}), which includes depth-averaged equations in the open water and cavity.
In the open water ($x<0$)
\begin{subequations}\label{eqs:sma_open_raw}
\begin{align}
a_{\tn{op}}\,(\varphi''+k^{2}\,\varphi)
+ 
\tanh(k\,H)\,\{\varphi - \eta_{\tn{w-a}}\} & = 0
\mathwhere
a_{\tn{op}}=\int_{-H}^{0}\frac{\cosh^{2}\{k\,(z+H)\}}{\cosh^{2}(k\,H)}\wrt{}z
\tag{\theequation a,b}
\\[8pt]
\mathandR 
k\,\tanh(k\,H)\,\varphi-\frac{\omega^{2}}{g}\,\eta_{\tn{w-a}}&=0
\quad\Rightarrow\quad
\varphi - \eta_{\tn{w-a}} = 0 
\quad\Leftrightarrow\quad
k\,\tanh(k\,H)=\frac{\omega^{2}}{g},
\tag{\theequation c,d,e}
\end{align}
\end{subequations}
so that $k\in\mathbb{R}^{+}$ used in Eq.~(\ref{eq:singlemodeapx_openwater}) satisfies the standard \textit{open water dispersion relation} (Fig.~\ref{fig:wavenumbers}).
Therefore, $\delta{}\mathcal{L}=0$ derives the \textit{field equation of the open water single-mode approximation:}
\begin{tcolorbox}[width=\linewidth, colback=white!95!blue]
\begin{equation}
\varphi''+k^{2}\,\varphi = 0
\mathfor x<0,
\end{equation}
\end{tcolorbox}
which has the general solution
\begin{equation}\label{eq:openocean_gs}
\varphi(x) = A^{(\tn{op})}\,\e^{\ci\,k\,x}+B^{(\tn{op})}\,\e^{-\ci\,k\,x},
\end{equation}
for as yet unspecified constants $A^{(\tn{op})}$ and $B^{(\tn{op})}$.

The depth-averaged equation in the cavity ($x>0$) is
\begin{subequations}\label{eqs:sma_cavity}
\begin{align}
	a_{\tn{ca}}\,\psi''
	+
	\{\kappa^{2}\,a_{\tn{ca}} -\kappa\,\tanh\{\kappa\,(H-d)\}\}\,\psi
	+ \frac{\omega^{2}}{g}\,w 
	= 0,	
	\\[10pt]
	\mathwhereR
	a_{\tn{ca}}=\int_{-H}^{-d}\frac{\cosh^{2}\{\kappa\,(z+H)\}}{\cosh^{2}\{\kappa\,(H-d)\}}\wrt{}z.
\end{align}
\end{subequations}
The remaining equations in the shelf--cavity involving the flexural shelf displacement are
\begin{subequations}\label{eqs:sma_w_raw}
\begin{align}
-\rho_{i}\,h\,\omega^{2}\,w + \frac{h^3\,\{(\lambda+2\,\mu)\,w'''' + \rho_{i}\,\omega^{2}\,w''\}}{12} 
	+ g\,\rho_{\tn{w}}\,(\eta_{\text{w--i}}-\psi) 
	+ g\,\rho_{i}\,(\eta_{\text{i--a}}-\eta_{\text{w--i}})	=0
	\tag{\theequation a}
\\[10pt]
w=\eta_{\tn{i-a}}
	\mathand
	(\rho_{\tn{i}}-\rho_{\tn{w}})\,(w-\eta_{\tn{w-i}})=0
	\quad
	\Rightarrow\quad
	w=\eta_{\tn{w-i}}=\eta_{\tn{i-a}}.
\tag{\theequation b,c,d}
\end{align}
\end{subequations}
Therefore, enforcing $\delta{}\mathcal{L}=0$ derives the \textit{coupled field equations of the single-mode and thin-plate approximations}:
\begin{tcolorbox}[width=\linewidth, colback=white!95!blue] 
\begin{subequations}\label{eqs:SMA_field}
\begin{align}
(1-m\,\omega^{2})\,w + F\,w'''' + J\,\omega^{2}\,w'' 
	- \psi 
	& =0
\\[10pt]
	a_{\tn{ca}}\,\psi''
	+
	\{\kappa^{2}\,a_{\tn{ca}} -\kappa\,\tanh\{\kappa\,(H-d)\}\}\,\psi
	+ \frac{\omega^{2}}{g}\,w 
	& = 0,
\\[10pt]
\mathandR
G\, u'' + m\,\omega^{2}\,u & = 0
\end{align}
\end{subequations}
\end{tcolorbox}
for $x>0$,
where
\begin{equation}
	F\equiv{}\frac{(\lambda+2\,\mu)\,h^{3}}{12\,\rho_{\tn{w}}\,g}
	,\quad
	G\equiv{} \frac{h\,(\lambda+2\,\mu)}{\rho_{\tn{w}}\,g}
	,\quad
	J\equiv{}\frac{\rho_{\tn{i}}\,h^{3}}{12\,\rho_{\tn{w}}\,g} 
	\mathand
	m \equiv \frac{\rho_{\tn{i}}\,h}{\rho_{\tn{w}}\,g}.
\end{equation}

Eqs.~(\ref{eqs:SMA_field}a,b) are identical to the single-mode approximation of \cite{porter2004approximations,bennetts_multi-mode_2007}, under the plane strain assumption $\lambda+2\,\mu \mapsto E\,/\,(1-\nu^{2})$, except for the appearance of rotational inertia.
Therefore, adapting \citet{porter2004approximations} and \citet{bennetts_multi-mode_2007} to include rotational inertia, the general solutions are 
\begin{subequations}\label{eqs:shelf_cavity_gs}
\begin{align}
\psi(x) & = A^{(\tn{ca})}\,\e^{\ci\,\kappa\,x}+B^{(\tn{ca})}\,\e^{-\ci\,\kappa\,x} 
+
\sum_{n=1,2}\Big\{A^{(\tn{ca})}_{-n}\,\e^{\ci\,\kappa_{-n}\,x}
+B^{(\tn{ca})}_{-n}\,\e^{-\ci\,\kappa_{n}\,x} \Big\}
\\[10pt]
\mathandR
w(x) & = A^{(\tn{fl})}\,\e^{\ci\,\kappa\,x}+B^{(\tn{fl})}\,\e^{-\ci\,\kappa\,x} 
+ \sum_{n=1,2}\Big\{A^{(\tn{fl})}_{-n}\,\e^{\ci\,\kappa_{-n}\,x}
+B^{(\tn{fl})}_{-n}\,\e^{-\ci\,\kappa_{-n}\,x} \Big\},
\end{align}
\end{subequations}
for as yet unspecified constants $A^{(\tn{ca})}$, $B^{(\tn{ca})}$, $A^{(\tn{fl})}$ and $B^{(\tn{fl})}$, such that
\begin{subequations}\label{eq:amp_relations}
\begin{align}
	A^{(\tn{ca})} & = \frac{\omega^{2}}{g\,\kappa\,\tanh\{\kappa\,(H-d)\}}\,A^{(\tn{fl})}
	\\[10pt]
	\mathandR
	A^{(\tn{ca})}_{-n} & = a_{\tn{ca}}^{-1}\,\{F\,(\kappa^{2}+\kappa_{-n}^{2})-J\,\omega^{2}\}\,\kappa\,\tanh\{\kappa\,(H-d)\}\,A^{(\tn{fl})}_{-n} \quad(n=1,2),
\end{align}
\end{subequations}
and similarly for the constants related to the left-going waves.
The wavenumber $\kappa$ is a root of the \textit{flexural-gravity wave dispersion equation}
\begin{equation}
\{F\,\kappa^{4}-J\,\omega^{2}\,\kappa^{2}+1-m\,\omega^{2}\}\,\kappa\,\tanh\{\kappa\,(H-d)\} = \frac{\omega^{2}}{g}.
\end{equation}
For low frequencies, the flexural-gravity wavenumber, $\kappa$, is similar to the open-water wavenumber, $k$, as restoring due to flexure (and rotational inertia) are negligible, but slightly larger due to the reduced water depth, i.e., $H-d<H$ (Fig.~\ref{fig:wavenumbers}).
For high frequencies, the flexural restoring dominantes and the flexural-gravity wavenumber becomes much smaller than the open water wavenumber. 
The wavenumbers $\kappa_{-n}\in\mathbb{R}+\ci\mathbb{R}^{+}$ ($n=1,2$) are roots of the quartic equation
\begin{equation}
a_{\tn{ca}}\,(F\,\kappa_{-n}^{4} - J\,\omega^{2}\,\kappa^{2}_{-n}  + 1 -m\,\omega^{2})
+
\{F\,(\kappa^{2}+\kappa_{-n}^{2})- J\,\omega^{2}\}\,\kappa\,\tanh\{\kappa\,(H-d)\}\}=0,
\end{equation}
which typically satisfy $\kappa_{-2}=-\kappa_{1}^{\ast}$, where $^{\ast}$ denotes the complex conjugate \cite{williams2006reflections,bennetts2007wave}.

Eq.~(\ref{eqs:SMA_field}c) for the extensional component of the shelf motions has the general solution
\begin{equation}
u(x) = A^{(\tn{ex})}\,\e^{\ci\,q\,x}+B^{(\tn{ex})}\,\e^{-\ci\,q\,x},
\end{equation}
for as yet unspecified constants $A^{(\tn{ex})}$ and $B^{(\tn{ex})}$.
The extensional wavenumber, $q$, is 
\begin{equation}
q = \omega\,\sqrt{\frac{m}{G}},	
\end{equation}
which is typically much smaller than the flexural-gravity wavenumber (and the open water wavenumber; Fig.~\ref{fig:wavenumbers}).

\begin{figure}
\centering
        \setlength{\figurewidth}{0.8\textwidth} 
        \setlength{\figureheight}{0.5\textwidth} 
        \input{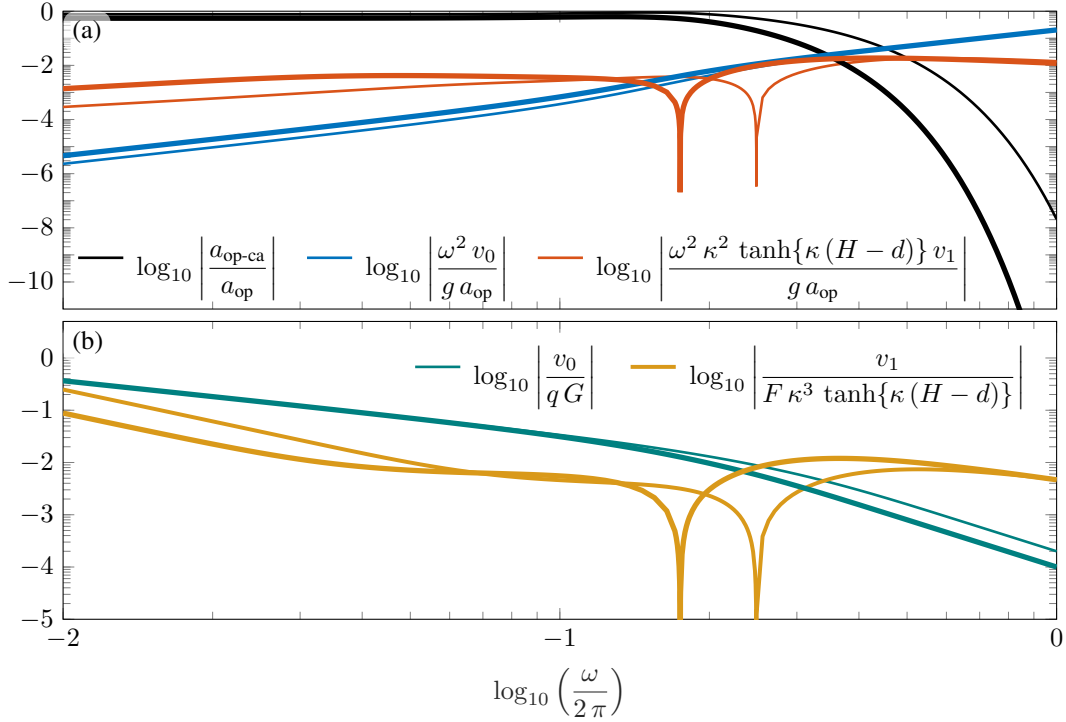} 
\caption{Normalised coefficients of the (a)~jump condition Eq.~(\ref{eqs:jumps}b) and (b)~shelf edge conditions Eqs.~(\ref{eqs:shelf_front_conds}a,b), which couple the open water to the shelf, versus frequency, for ice thickness $h=200$\,m (thin curves) and $h=400$\,m (thick) and water depth $H=800$\,m.}	
\label{fig:coupling_coeffs}
\end{figure}

The contribution to $\delta{}\mathcal{L}$ on the interface between the open ocean and the cavity is
\begin{align} \nonumber
\delta\mathcal{C}_{\tn{op-ca}} = &	
	-\rho_{\tn{w}}\,g\,\int_{t_{0}}^{t_{1}} \e^{-2\,\ci\,\omega\,t}\,[\delta{}\varphi]_{x=0}\Bigg\{
	\int_{-H}^{0} 
	     [\varphi']_{x=0}\frac{\cosh^{2}\{k\,(z+H)\}}{\cosh^{2}(k\,H)}\wrt{}z
	\\[6pt] \nonumber & ~~~~~~~~~~~~~~~~~~~
	- \int_{-H}^{-d} 
	     [\psi']_{x=0}\frac{\cosh\{k\,(z+H)\,\cosh\{\kappa\,(z+H)\}}{\cosh(k\,H)\cosh\{\kappa\,(H-d)\}}\wrt{}z
	\\[6pt] \nonumber & ~~~~~~~~~~~~~~~~~~~     
	 - \int_{-d}^{0}\frac{\omega^{2}}{g}\,\frac{\cosh\{k\,(z+H)\}}{\cosh\{\kappa\,(H-d)\}} \,\Big\{u-\Big(d-\frac{h}{2}+z\Big)\,w' \Big\}\wrt{}z\Bigg\}\wrt{}t
\\[10pt] \nonumber & +
	\int_{t_{0}}^{t_{1}} \e^{-2\,\ci\,\omega\,t}\,[\delta{}\psi']_{x=0}\Bigg\{
	\int_{-H}^{0} 
	[\psi]_{x=0}\frac{\cosh^{2}\{\kappa\,(z+H)\}}{\cosh^{2}\{\kappa\,(H-d)\}}\wrt{}z
	\\[6pt] & ~~~~~~~~~~~~~~~~~~~
	- \int_{-H}^{-d} 
	     [\psi']_{x=0}\frac{\cosh\{k\,(z+H)\,\cosh\{\kappa\,(z+H)\}}{\cosh(k\,H)\cosh\{\kappa\,(H-d)\}}\wrt{}z\Bigg\}\wrt{}t.
\end{align}
Setting $\delta\mathcal{C}_{\tn{op-ca}}=0$ leads to the \textit{interfacial ``jump'' conditions for single-mode approximation:}
\begin{tcolorbox}[width=\linewidth, colback=white!95!blue] 
\begin{subequations}\label{eqs:jumps}
\begin{gather}
	a_{\tn{op-ca}}\,\varphi
	=
	a_{\tn{ca}}\,\psi
	\mathand
	a_{\tn{op}}\,\varphi'  = 
	a_{\tn{op-ca}}\,\psi' 
	+ \frac{\omega^{2}}{g}\,\Big\{v_{0}\,u - v_{1}\,w'\Big\}
	\tag{\theequation a,b}
\end{gather}
\end{subequations}
\end{tcolorbox}
for $x=0$,
where
\begin{align}
	a_{\tn{op-ca}}& 
	=
	\int_{-H}^{-d}
	\frac{\cosh\{k\,(z+H)\}\,\cosh\{\kappa\,(z+H)\}}{\cosh(k\,H)\,\cosh\{\kappa\,(H-d)\}}
	~\textrm{d}z,
\\[10pt]
	v_{0} & =
	\int_{-d}^{0}
	\frac{\cosh\{k\,(z+H)\}}{\cosh(k\,H)}~\textrm{d}z
\\[10pt]
	\mathandR
	v_{1} & =
	\int_{-d}^{0}
	\frac{\cosh\{k\,(z+H)\}}{\cosh(k\,H)}\left(d-\frac{h}{2}+z\right)~\textrm{d}z.
\end{align}
\new{Eq.~(\ref{eqs:jumps}a) is a weak form of continuity of pressure between the open ocean and sub-shelf water cavity.
Eq.~(\ref{eqs:jumps}b) is a weak form of continuity of horizontal water velocity between the open ocean and combined water and shelf front.
The jump conditions} are identical to the jump conditions derived by \cite{porter2004approximations,bennetts_multi-mode_2007} (restricted to piecewise constant geometry), except that the integration of the coefficient $v_{\tn{op}}$ extends to the free surface ($z=0$) rather than the ice underside ($z=-d$), and as a result of the appearance of the ice displacements in Eq.~(\ref{eqs:jumps}b).
For low frequencies, the (normalised) coefficient of the cavity water velocity in Eq.~(\ref{eqs:jumps}b) is much greater than the (normalised) coefficients of the ice displacement/velocity (Fig.~\ref{fig:coupling_coeffs}a), indicating that the jump condition is dominated by the depth averaged water velocities.
The coefficients of the ice displacement/velocity increase with frequency, whereas the coefficient of water velocity decreases, such that the former become comparable and then much greater than the latter, which indicates the jump condition provides strong coupling between the open ocean and shelf.

The contribution to $\delta{}\mathcal{L}$ on the interface between the open ocean and the ice shelf is
\begin{align}\nonumber
\delta\mathcal{C}_{\tn{op-sh}} = &
	\int_{t_{0}}^{t_{1}}
	[\delta{u}]_{x=0}\,\Bigg\{\e^{-2\,\ci\,\omega\,t}\, 
	 \Big(
	 h\,(\lambda+2\,\mu)\, [u']_{x=0} 
	 + 
	 \rho_{\tn{w}}\,g\,[\varphi]_{x=0}\,\int_{-d}^{0}\frac{\cosh\{k\,(z+H)\}}{\cosh(k\,H)}\wrt{}z\Big)
	\\[6pt] \nonumber
	& ~~~~~~~~~~~~~~~~~~~~~~~~
	+ \e^{-\ci\,\omega\,t}\,\Big(
	\int_{-d}^{0}\,P_{\tn{at}}-\rho_{\tn{w}}\,g\,z\wrt{}z
	\int_{-d}^{0}\,P_{\tn{at}}\wrt{}z\Big)
	\Bigg\} ~\textrm{d}t
\\[10pt] \nonumber
	& ~ 
	+ 
	\int_{t_{0}}^{t_{1}}
	\e^{-2\,\ci\,\omega\,t}\,
	 [\delta{}w]_{x=0}\,
	\Big(\frac{-\rho_{i}\,h^{3}\,\omega^{2}}{12}\,[w']_{x=0} - \frac{h^{3}\,(\lambda+2\,\mu) }{12}\,[w''']_{x=0} \Big)
	~\textrm{d}t
\\[10pt]\nonumber 
	& ~ 
	+ \int_{t_{0}}^{t_{1}}
	[\delta{}w']_{x=0}\,
	\Bigg\{ 
	\e^{-2\,\ci\,\omega\,t}\,
	\Bigg(\frac{h^{3}\,(\lambda+2\,\mu)}{12}\,[w'']_{x=0} 
	\\[6pt]\nonumber 
	& ~~~~~~~~~~~~~~~~~~~~~~~~~~~~~~~~~~~~~~~~~~~~~~~
	- 
	\rho_{\tn{w}}\,g\,[\varphi]_{x=0}\,\int_{-d}^{0}
	\left(d-\frac{h}{2}+z\right)\,\frac{\cosh\{k\,(z+H)\}}{\cosh(k\,H)}~\wrt{}z \Bigg)
	\\[6pt] & ~~~~~~~~~~~~~~~~~~~~~~~~~~~~
	+ \e^{-\ci\,\omega\,t}\,\Big(
	\int_{-d}^{0}\left(d-\frac{h}{2}+z\right)\,\left(P_{\tn{at}}-\rho_{\tn{w}}\,g\,z\right)\wrt{}z
	\int_{-d}^{0}\left(d-\frac{h}{2}+z\right)\,P_{\tn{at}}\wrt{}z\Big)
	\Bigg\} ~\textrm{d}t.
\end{align}
Setting $\delta\mathcal{C}_{\tn{op-sh}} =0$ leads to the \textit{(dynamic, $\omega\neq{}0$) shelf front conditions}:
\begin{tcolorbox}[width=\linewidth, colback=white!95!blue,] 
\begin{subequations}\label{eqs:shelf_front_conds}
\begin{equation}
	G\, u' + v_{0}\,\varphi = 0,
	\quad
	F\,w'' + v_{1}\,\varphi = 0
	\mathand
	F\, w''' + J\,\omega^{2}\,w' =0
	\tag{\theequation a,b,c}
\end{equation}
\end{subequations}
\end{tcolorbox}
for $x=0$.
Eqs.~(\ref{eqs:shelf_front_conds}a,b) couple the ice and open water displacements. 
The ratios of the coefficients in the coupling conditions (Fig.~\ref{fig:coupling_coeffs}b) indicate (i)~strong coupling in Eq.~(\ref{eqs:shelf_front_conds}a) at low frequencies, degenerating to uncoupled zero normal traction at higher frequencies, and (ii)~Eq.~(\ref{eqs:shelf_front_conds}b) is approximately the bending moment component of the standard (uncoupled) free edge conditions over the frequency range considered.


\subsection{Scattering matrix} 

The jump conditions (\ref{eqs:jumps}) and shelf front conditions (\ref{eqs:shelf_front_conds}) are applied to the general solutions  (\ref{eq:openocean_gs}) and (\ref{eqs:shelf_cavity_gs}) to derive a system of relations between the amplitudes of the waves that propagate/decay towards and away from $x=0$, $A^{(\bullet)}$ and $B^{(\bullet)}$, respectively.
Restricting to propagating waves only, and using Eq.~(\ref{eq:amp_relations}) to eliminate $A^{(\tn{ca})}$ and $B^{(\tn{ca})}$, derives the scattering matrix, $\mathcal{S}$, which relates the outgoing amplitudes to the incoming amplitudes, such that 
\begin{equation}
	\left(\begin{array}{c}	
	B^{\tn{(op)}}
	\\
	B^{\tn{(fl)}}
	\\
	B^{\tn{(ex)}}
	\end{array}\right)
	=
	\mathcal{S}\,
	\left(\begin{array}{c}	
	A^{\tn{(op)}}
	\\
	A^{\tn{(fl)}}
	\\
	A^{\tn{(ex)}}
	\end{array}\right)
	\mathwhere
	\mathcal{S}=
	\left(
	\begin{array}{ccc}
		\mathcal{R}^{\tn{(op$\rightarrow$op)}} 
		& \mathcal{T}^{\tn{(fl$\rightarrow$op)}} 
		& \mathcal{T}^{\tn{(ex$\rightarrow$op)}}
		\\
		\mathcal{T}^{\tn{(op$\rightarrow$fl)}} 
		& \mathcal{R}^{\tn{(fl$\rightarrow$fl)}} 
		& \mathcal{R}^{\tn{(ex$\rightarrow$fl)}}
		\\
		\mathcal{T}^{\tn{(op$\rightarrow$ex)}} 
		& \mathcal{R}^{\tn{(fl$\rightarrow$ex)}} 
		& \mathcal{R}^{\tn{(ex$\rightarrow$ex)}}
	\end{array}
	\right),
	\tag{\theequation a,b}
\end{equation}
in which the $\mathcal{R}^{\bullet}$ and $\mathcal{T}^{\bullet}$ are, respectively, reflection and transmission coefficients to be found from the solution of the problem in \S\,\ref{sec:sma}.
\new{In general, $\mathcal{T}^{\tn{(op$\rightarrow$ex)}}\neq\mathcal{T}^{\tn{(ex$\rightarrow$op)}}$, etc, as $\mathcal{T}^{\tn{(op$\rightarrow$ex)}}$ is the coefficient of the extensional wave in the ice shelf forced by a unit-amplitude incident wave from the open ocean, whereas $\mathcal{T}^{\tn{(ex$\rightarrow$op)}}$ denotes the amplitude of a wave transmitted into the open ocean by an incident extensional wave from the ice shelf.
The latter is typically not a physical problem considered in wave--shelf interaction studies.}
Using standard methods \cite{porter2004approximations}, it can be deduced that 
\begin{equation}\label{eq:scat-id}
\mathcal{S}\,\mathcal{S}^{\ast}=\mathcal{I},	
\end{equation}
where $^{\ast}$ denotes the conjugate matrix and $\mathcal{I}$ is the $3\times{}3$ identity matrix, from which energy balances can be derived (see below).


\section{Results}\label{sec:results}

\begin{figure}
\centering
  \includegraphics[width=\textwidth]{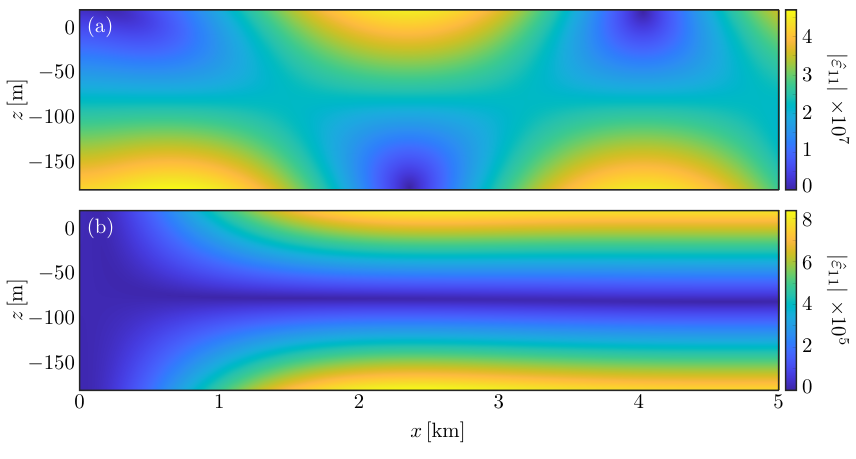}
  \caption{Strain fields up to 5\,km from the shelf front, for ice thickness $h=200$\,m, water depth $H=800$\,m, and wave period (a)~$T=15$\,s and (b)~$T=50$\,s.}	
  \label{fig:strain_fields}
\end{figure}

Consider the problem in which motions are excited by an ambient incident wave from the ocean ($A^{(\tn{fl})}=A^{(\tn{ex})}\equiv{}0$) at a prescribed period $T=2\,\pi\,/\,\omega$. 
Without loss of generality, a unit incident wave amplitude is set ($A^{(\tn{op})}=1$\,m).
The primary quantity of interest is the spatial component of the (non-zero) strain component
\begin{equation}
\hat{\varepsilon}_{11}(x,z:T)=u'-(z+d-h\,/\,2)\,w'',	
\end{equation}
which is such that $\varepsilon_{11}(x,z,t)=\hat{\varepsilon}_{11}(x,z)\,\e^{-\ci\,\omega\,t}$.
Examples of the strain field (Fig.~\ref{fig:strain_fields}) indicate that the extensional and flexural motions both contribute to the strain for relatively short periods (in the swell regime), as it has nonlinear structure in both spatial dimensions, whereas only the flexural motion contributes for longer periods (IG wave regime and above), indicated by the vertical symmetry about the unstrained mid-plane ($z=h\,/\,2-d$). 
The shelf front experiences strains comparable to the shelf interior for the shorter period and near-zero  strain for the longer period, where the latter is ensured by the exponentially decaying components of the flexural motion (with wavenumbers $\kappa_{-n}$ in Eq.~\ref{eqs:shelf_cavity_gs}b).

\begin{figure}[h!]
        \centering
        \setlength{\figurewidth}{0.85\textwidth} 
        \setlength{\figureheight}{0.5\textwidth} 
        \def\labelsA{1}
        \def\labelsB{1}
        \input{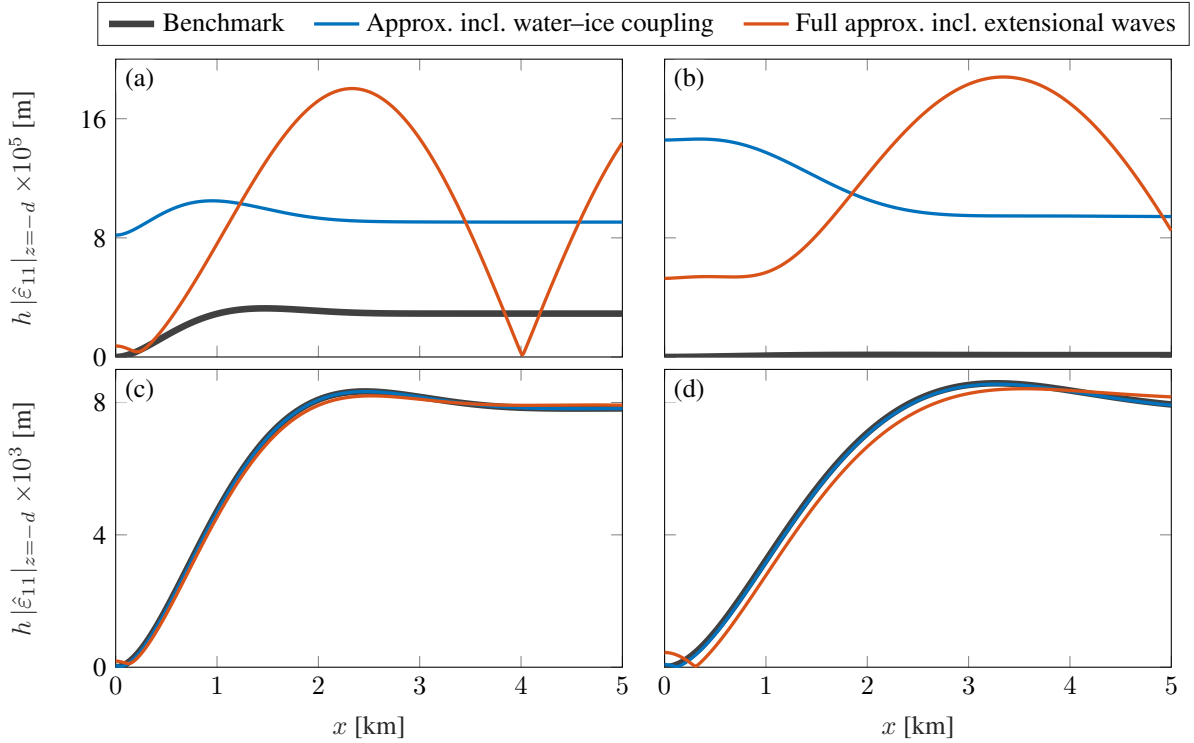} 
        \caption{Comparison of scaled strain profiles predicted by three models: (i)~the benchmark model without water--ice coupling at the shelf front and extensional waves \cite{porter2004approximations,bennetts_multi-mode_2007}; (ii)~the model in which water--ice coupling occurs at the shelf front through the velocity jump condition (Eq.~\ref{eqs:jumps}b) and the bending moment condition (Eq.~\ref{eqs:shelf_front_conds}b); and (iii)~the full model proposed in \S\,\ref{sec:frequencydomain} including extensional wave motion and water--ice coupling at the shelf front,  
        for shelf thickness (a,c)~$h=200$\,m and (b,d)~$h=400$\,m, and bed depth $H=800$\,m, in response to incident waves with period (a,b)~$T=15$\,s and (c,d)~$T=50$\,s.}
        \label{fig:strain_profiles}
\end{figure}

Example strain profiles at the lower ice shelf surface (Fig.~\ref{fig:strain_profiles}) show the influence of the additional terms in the thin-plate approximation. 
The strains are scaled by the shelf thickness, such that strains for different thickness values are of the same order of magnitude for the different wave periods.
The benchmark model (without water--ice coupling at the shelf front and extensional waves) predicts the strain modulus increases from zero at the shelf front to a maximum value after several kilometres, followed by a plateau at a slightly smaller value.

For the shorter wave period (Fig.~\ref{fig:strain_profiles}a,b), the addition of water--ice coupling at the shelf front (through Eqs.~\ref{eqs:jumps}b and \ref{eqs:shelf_front_conds}b) causes a large relative increase in the strain, by factors of $\approx{}3$ for the thinner shelf and $\approx{}75$ for the thicker shelf at the plateaus (approximately $x>3$\,km).
The strain at the shelf front is non-zero and, for the thicker shelf (Fig.~\ref{fig:strain_profiles}b), the greatest strain occurs at the shelf front, such that the wave--ice coupling causes a qualitative change in the strain profile.
The effect of wave--ice coupling on the strain profiles is almost imperceptible for the longer wave period (Fig.~\ref{fig:strain_profiles}c,d), although the strains are one to two orders of magnitude larger than for the shorter period, and the change in scale masks the similarity shelf front strain values for the respective thicknesses, as anticipated by the coupling coefficient in the bending moment condition (Fig.~\ref{fig:coupling_coeffs}b; yellow curves).

The addition of extensional waves changes the qualitative behaviour of the strain profiles for the shorter wave period (Fig.~\ref{fig:strain_profiles}a,b).
Notably, the strain oscillates away from the shelf front rather than reaching a plateau, due to interference between the flexural and extensional waves. 
The extensional waves have a far smaller effect on the strain profiles for the longer wave period (Fig.~\ref{fig:strain_profiles}c,d), although their influence for the thicker shelf (Fig.~\ref{fig:strain_profiles}d) is greater than that of the wave--ice coupling on the flexural waves.

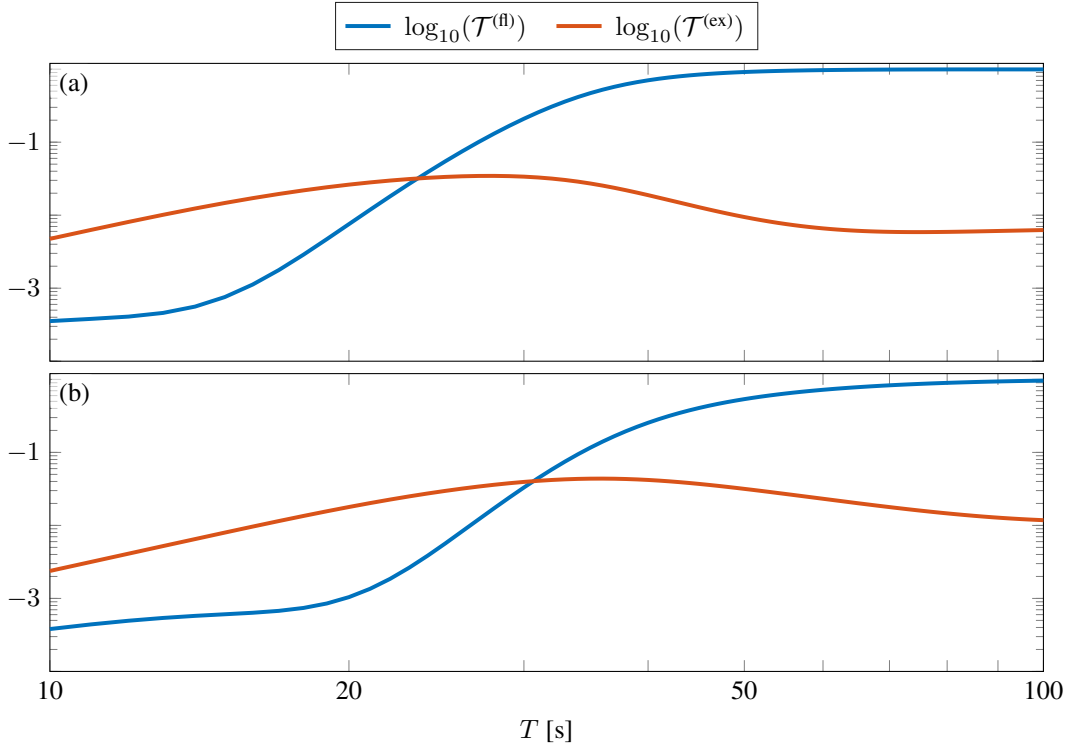
\begin{figure}[h!]
        \centering
        \setlength{\figurewidth}{0.8\textwidth} 
        \setlength{\figureheight}{0.5\textwidth} 
        \def\labelsA{1}
        \def\labelsB{1}

%
\begin{tikzpicture}

\begin{axis}[%
width=\figurewidth,
height=0.48\figureheight,
at={(0\figurewidth,0.5\figureheight)},
scale only axis,
xmode=log,
xmin=10,
xmax=100,
xminorticks=true,
xtick={10,20,...,100},
xticklabels={\empty},
ymode=log,
ymin=0.0001,
ymax=1.2,
yminorticks=true,
ytick={1e-4,1e-3,1e-2,1e-1,1e0},
yticklabels={$$,$-3$,$$,$-1$,$$},
axis background/.style={fill=white},
legend style={
at={(current axis.north)},
legend columns=-1,
/tikz/every even column/.append style={column sep=3mm},
anchor=south,
yshift=5pt,
draw=white!15!black}
]
\addplot [color=matlabblue, line width=1.5pt]
  table[row sep=crcr]{%
10	0.000354248461018135\\
11	0.000379983487487032\\
12	0.000408396467506614\\
13	0.000458671206968618\\
14	0.000559847523111857\\
15	0.000756853293781205\\
16	0.00112228490892389\\
17	0.0017719253584051\\
18	0.00288069653377811\\
19	0.00469620366513329\\
21	0.0118526500186128\\
22	0.0181109660000521\\
23	0.0269010396803664\\
24	0.0388599891460947\\
25	0.0546540867127154\\
26	0.0749327861341636\\
27	0.100264730072205\\
28	0.131057672402343\\
29	0.167470817501804\\
30	0.209335745974939\\
31	0.256107580561641\\
32	0.306867181334543\\
33	0.360385333707533\\
34	0.415242817151459\\
35	0.469982545054553\\
36	0.523259986692388\\
37	0.573960575921116\\
38	0.621265823897519\\
39	0.664666424924914\\
40	0.703933607326634\\
41	0.739065779669444\\
42	0.77022675295282\\
43	0.797687329036178\\
44	0.821776736409499\\
45	0.842846059130365\\
46	0.861243043799513\\
47	0.877296328745054\\
48	0.89130677778868\\
49	0.903543781116957\\
50	0.914244790789995\\
51	0.92361680076636\\
52	0.931838870906674\\
53	0.939065100987176\\
54	0.945427684974115\\
55	0.951039831082391\\
56	0.955998435818397\\
58	0.964275034952559\\
60	0.970789511860688\\
62	0.9759356700866\\
64	0.980007968593132\\
67	0.984575691418031\\
70	0.9877634949939\\
73	0.989948716989472\\
77	0.991754217984356\\
82	0.992821671662653\\
89	0.99303789641378\\
100	0.992135472008193\\
};
\addlegendentry{~$\log_{10}(\mathcal{T}^{\tn{(fl)}})$}
\addplot [color=matlabred, line width=1.5pt]
  table[row sep=crcr]{%
10	0.00474452978646148\\
11	0.00628731450377311\\
12	0.00808946777180991\\
13	0.0101254483757797\\
14	0.0123496084732292\\
15	0.0147022451363579\\
16	0.0171174957735948\\
17	0.0195305233557092\\
18	0.0218825661006191\\
19	0.0241235473421001\\
20	0.02621263235115\\
21	0.028117368765602\\
22	0.0298120234235383\\
23	0.031275611709947\\
24	0.0324900208992224\\
25	0.0334385864367479\\
26	0.0341054563908829\\
27	0.0344760162900269\\
28	0.0345384941023139\\
29	0.0342866032297981\\
30	0.0337227401716228\\
31	0.0328609300605462\\
32	0.0317285614923113\\
33	0.0303661169399068\\
34	0.028824614552119\\
35	0.027161170022885\\
36	0.0254336659682446\\
37	0.0236957202209007\\
38	0.0219929144500765\\
39	0.0203607426431406\\
40	0.0188242197567529\\
42	0.0160916904859501\\
45	0.0128726009428185\\
46	0.0120145746762322\\
47	0.0112506540472732\\
48	0.0105721564603535\\
49	0.00997063541005236\\
50	0.00943812219633856\\
51	0.00896725449680411\\
52	0.00855132843750409\\
53	0.00818430138895764\\
54	0.00786076486680658\\
55	0.00757590083360129\\
56	0.0073254302097592\\
57	0.00710555920990233\\
58	0.00691292691858337\\
59	0.00674455604080637\\
60	0.00659780779678394\\
61	0.0064703413185712\\
62	0.00636007753381091\\
63	0.00626516730790827\\
64	0.00618396350467739\\
65	0.00611499657898059\\
66	0.00605695330786994\\
67	0.00600865828263117\\
68	0.00596905781225888\\
69	0.00593720592251193\\
70	0.00591225216972721\\
71	0.00589343102255133\\
72	0.00588005259641555\\
73	0.00587149455430137\\
74	0.00586719501291367\\
75	0.00586664631585727\\
76	0.00586938955500116\\
78	0.00588313151588844\\
80	0.00590555435722225\\
82	0.00593431423389537\\
85	0.00598525103401034\\
89	0.00606032621345275\\
98.0000000000001	0.00622497139178689\\
100	0.00625730458780033\\
};
\addlegendentry{~$\log_{10}(\mathcal{T}^{\tn{(ex)}})$}
\node[align=center,fill=white,opacity=0.6,text opacity=1,rounded corners,anchor=north west]
at (current axis.north west) {(a)};
\end{axis}

\begin{axis}[%
width=\figurewidth,
height=0.48\figureheight,
at={(0\figurewidth,0\figureheight)},
scale only axis,
xmode=log,
xmin=10,
xmax=100,
xminorticks=true,
scaled x ticks=false,
xlabel={$T$ [s]},
xtick={10,20,...,100},
xticklabels={10,20,,,50,,,,,100},
ymode=log,
ymin=0.0001,
ymax=1.2,
ytick={1e-4,1e-3,1e-2,1e-1,1e0},
yticklabels={$$,$-3$,$$,$-1$,$$},
yminorticks=true,
axis background/.style={fill=white}
]
\addplot [color=matlabblue, line width=1.5pt, forget plot]
  table[row sep=crcr]{%
10	0.000380744437162578\\
11	0.000440816975882845\\
12	0.000494856159703134\\
13	0.000540452706979238\\
14	0.000576976167056639\\
16	0.000636209939349179\\
17	0.000675773481169492\\
18	0.000740787086473408\\
19	0.000852599468101389\\
20	0.00104175394900546\\
21	0.00135331014315169\\
22	0.00185409369362453\\
23	0.00264104518825285\\
24	0.00384941714363884\\
25	0.0056592747562626\\
27	0.0120408725691819\\
28	0.0171964321534572\\
29	0.0240951010404289\\
30	0.0330618801830288\\
31	0.0443867971451129\\
32	0.0582930720997195\\
33	0.0749087251119833\\
34	0.0942469182794654\\
35	0.116198971051027\\
36	0.140541436323191\\
37	0.166955680692396\\
38	0.195056032073494\\
39	0.224421415814331\\
40	0.254625632761007\\
41	0.285262684112691\\
42	0.315965228928957\\
43	0.346415829779127\\
44	0.376351774629373\\
45	0.4055648704676\\
46	0.433897764409908\\
47	0.461238210793548\\
48	0.487512417062729\\
49	0.512678278908734\\
50	0.53671902295744\\
51	0.559637543004988\\
52	0.581451548662658\\
53	0.602189535601045\\
54	0.621887521786475\\
55	0.640586461348485\\
56	0.658330236119383\\
57	0.675164126151569\\
58	0.69113366880228\\
59	0.706283827430489\\
60	0.720658403007028\\
61	0.734299633653822\\
62	0.747247937629837\\
63	0.759541764314422\\
64	0.771217525288428\\
65	0.782309583791632\\
67	0.802870020006942\\
69	0.821458839445337\\
71	0.838283267061519\\
73	0.853525096455614\\
75	0.867343830281522\\
77	0.879879621255484\\
79	0.891255914842138\\
81	0.901581766215787\\
83.0000000000001	0.910953841164913\\
85	0.919458128049104\\
88	0.930753171719475\\
91.0000000000001	0.940493058987224\\
94	0.948871361418354\\
97	0.956054716798367\\
100	0.962187535234423\\
};
\addplot [color=matlabred, line width=1.5pt, forget plot]
  table[row sep=crcr]{%
10	0.00237844556408099\\
13	0.00521562631696729\\
15	0.00797825052009196\\
16	0.0096378704293721\\
17	0.0114760649127943\\
18	0.0134778660122526\\
19	0.0156210705420037\\
20	0.0178777055489651\\
21	0.0202159148663135\\
22	0.0226018426775984\\
23	0.0250011680910248\\
24	0.0273800886961268\\
25	0.0297057274354579\\
26	0.0319461062404718\\
27	0.0340699501870865\\
28	0.0360466274586867\\
29	0.0378464841599673\\
30	0.0394417113987039\\
31	0.0408077149913595\\
32	0.0419247886114993\\
33	0.0427797676453525\\
34	0.0433673037953975\\
35	0.0436904651618937\\
36	0.0437605139006639\\
37	0.0435958949123764\\
38	0.043220626428562\\
39	0.0426623738055458\\
40	0.0419504979518002\\
41	0.0411143148789097\\
42	0.0401817139384335\\
43	0.039178190770003\\
44	0.0381262782285109\\
45	0.0370453131878411\\
46	0.0359514576925497\\
47	0.0348578927068391\\
48	0.033775113790638\\
49	0.0327112738054249\\
51	0.030663396867659\\
53	0.0287454359189492\\
56	0.0261383153201567\\
65	0.0201607046046814\\
68	0.0186867931663552\\
71	0.017420750391638\\
73	0.0166788216144959\\
75	0.0160102064722281\\
77	0.0154084720358493\\
79	0.014867788962765\\
81	0.0143828770473944\\
83.0000000000001	0.0139489506930526\\
85	0.0135616672288102\\
87	0.0132170794663637\\
89	0.0129115930000143\\
91.0000000000001	0.0126419282501422\\
93	0.0124050869834146\\
95.0000000000001	0.0121983229163498\\
97	0.0120191159642231\\
99	0.0118651496988982\\
100	0.0117969587173055\\
};
\node[align=center,fill=white,opacity=0.6,text opacity=1,rounded corners,anchor=north west]
at (current axis.north west) {(b)};
\end{axis}
\end{tikzpicture}%
        \caption{Transmitted energy proportions for flexural and extensional waves versus wave period, for shelf thickness (a)~$h=200$\,m and (b)~$h=400$\,m, and bed depth $H=800$\,m.}
        \label{fig:transmission}
\end{figure}

The proportion of incident wave energy transmitted into the flexural and extensional waves is used to assess their relative influence on the ice shelf motion versus wave period.  
The distribution of incident wave energy is derived from Eq.~(\ref{eq:scat-id}), which gives the energy balance
\begin{equation}\label{eq:energy-id}
	\mathcal{R}+\mathcal{T}^{\tn{(fl)}}+\mathcal{T}^{\tn{(ex)}}=1,
\end{equation}
where 
\begin{equation}
	\mathcal{R}=\vert{}\mathcal{R}^{\tn{(op$\rightarrow$op)}}\vert^{2}
	,\quad
	\mathcal{T}^{\tn{(fl)}}=\vert{}\mathcal{T}^{\tn{(fx$\rightarrow$op)}}\vert\,\vert{}\mathcal{T}^{\tn{(op$\rightarrow$fl)}}\vert
	\mathand
	\mathcal{T}^{\tn{(ex)}}=\vert{}\mathcal{T}^{\tn{(ex$\rightarrow$op)}}\vert\,\vert{}\mathcal{T}^{\tn{(op$\rightarrow$ex)}}\vert.
\end{equation}
are the proportions of the incident energy in the reflected wave ($\mathcal{R}$), and the flexural ($\mathcal{T}^{\tn{(fl)}}$) and extensional ($\mathcal{T}^{\tn{(ex)}}$) waves transmitted into the shelf--cavity region.

For periods in the majority of the swell regime, the transmitted extensional waves carry more energy than the flexural waves (Fig.~\ref{fig:transmission}).
The difference is approximately an order of magnitude for the shortest periods considered, and is greatest for the thinner shelf (Fig.~\ref{fig:transmission}a).
The proportion of energy in the flexural waves increases steeply as wave period transitions from the swell to infragravity regimes, whereas the proportion of energy in the extensional waves slightly decreases.
This causes the flexural wave energy to exceed the extensional wave energy in the infragravity wave regime, with the difference approximately two orders of magnitude at the longest wave periods considered and greater for the thinner shelf.
The wave period at which the energies of the flexural and extensional waves are equal is longer for the thicker shelf than the thinner shelf ($\approx{}30$\,s vs.\ $\approx{}23$\,s).
 
\begin{figure}[h!]
        \centering
        \setlength{\figurewidth}{0.8\textwidth} 
        \setlength{\figureheight}{0.5\textwidth} 
        \def\labelsA{1}
        \def\labelsB{1}
        \input{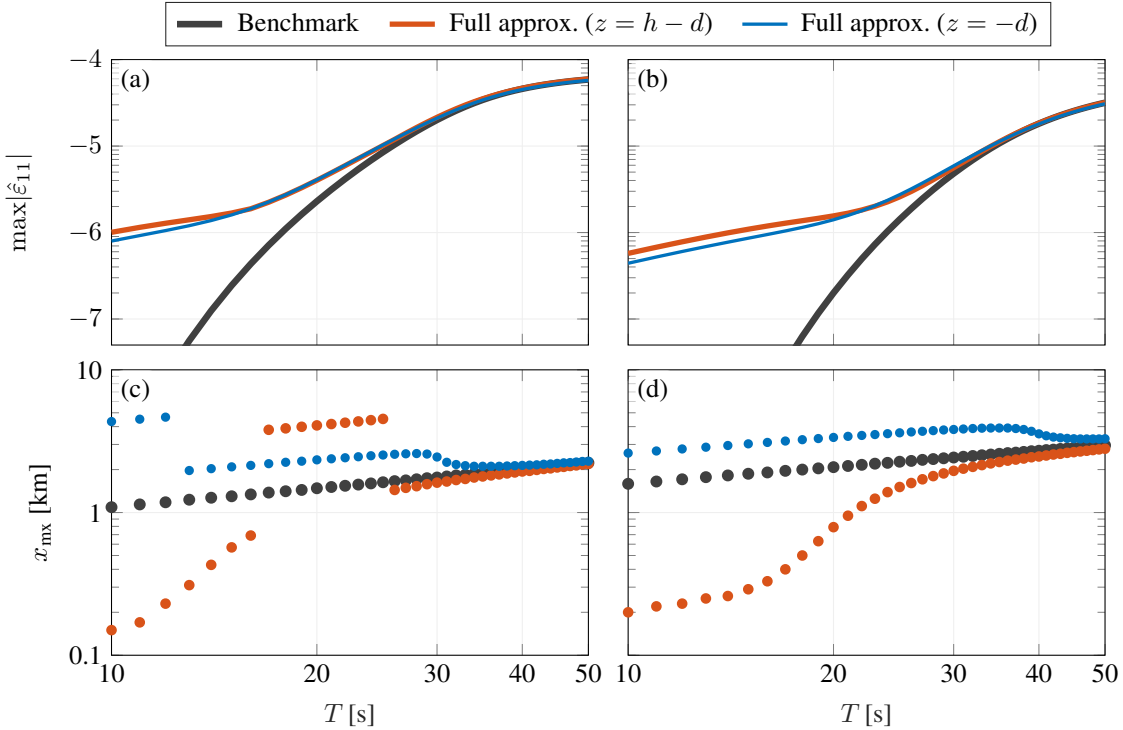} 
        \caption{(a,b)~Maximum flexural strains at upper ($z=h-d$) and lower ($z=-d$) shelf surfaces, and (c,d)~corresponding locations, for shelf thickness (a)~$h=200$\,m and (b)~$h=400$\,m, and bed depth $H=800$\,m, with results of benchmark model shown for reference.}
        \label{fig:maxstrains}
\end{figure}

For the cases tested with the full approximation outlined in \S{}\ref{sec:frequencydomain}, the maximum shelf strains are attained at either the upper or lower shelf surface, which is similar to the benchmark model, where the  maximum strains are attained at both upper and lower surfaces due to symmetry about the mid-plane.
In the swell regime, the maximum strains predicted by the full model far exceed those of the benchmark model (Fig.~\ref{fig:maxstrains}a,b).
The maximum strains at the upper surfaces slightly exceed those at the lower surface for the smallest wave periods considered.
For longer periods, the maximum strains at the upper and lower surfaces are almost identical, and they tends towards the maximum strain predicted by the benchmark model, as the wave period increases, such that they are indistinguishable in the infragravity regime.

For the shortest wave period considered, the strain maxima at the upper surface occur only hundreds of metres from the shelf front, and move closer towards the shelf front as wave period decreases (Fig.~\ref{fig:maxstrains}c,d).
In contrast, the maxima predicted by the benchmark model occur more than a kilometre away from the shelf front, and the maxima at the lower surface predicted by the full approximation occur even farther away.
For shorter periods and the thinner shelf, the strain maxima move between distinct regions of large strain at the upper and lower surfaces (yellow patches in Fig.~\ref{fig:strain_fields}a), which causes the jumps in the locations of maximum strain (Fig.~\ref{fig:maxstrains}c).

\section{Conclusions and discussion}

The governing equations for the canonical problem of incident waves from the open ocean forcing motions of a floating ice shelf, in which the ice shelf is modelled by the full equations of elasticity and has an Archimedean draught, have been derived from a variational principle.
The variational principle was used to derive thin-plate equations for the ice shelf, involving both flexural and extensional waves, and a single-mode approximation in the water.
Results have shown that the water--ice coupling at the submerged portion of the shelf front and the extensional waves, which are neglected in the benchmark thin-plate model for floating ice, significantly increase shelf strains  for wave periods in the swell regime.
In contrast, they have a negligible effect for periods in the infragravity wave regime.


Variational principles are often used to derive approximations for water wave problems, dating back to \citet{luke1967variational}, and with the so-called (modified) mild-slope equations of \citet{miles1991variational}, \citet{chamberlain1995modified} and others of particular relevance to the present study.
The variational principle presented in \S\,\ref{sec:vp} can be viewed as an extension of the variation principle of \citet{porter2004approximations} to incorporate the full equations of elasticity for the floating ice.
However, there are notable differences in the approach used here, which is arguably more closely aligned to the `unified theory' of \citet{porter2020mild} for open water waves.
In particular, our use of a displacement potential in the water, for consistency with the unknown displacements in the ice, is a major departure from \citet{porter2004approximations} and others before.
Further, we include interfacial stresses in the variational principle, so that all matching conditions arise as natural conditions of the variational principle, and essential conditions do not have to be imposed as in \cite{porter2004approximations}.

\new{
There is evidence from studies on cognate problems (without water--ice coupling at the ice edge and extensional waves) that the single-mode approximation is accurate \cite{bennetts_multi-mode_2007,bennetts_interaction_2009,bennetts2021complex}.
In general, the single-mode approximation becomes less accurate as frequency increases and impedance mismatch increases (Fig.~\ref{fig:wavenumbers}).
In this case, following \citet{bennetts_multi-mode_2007}, the single-mode approximations (Eq.~\ref{eqs:SMA}), can be extended to include an arbitrary, finite number of vertical modes that support evanescent waves, such that continuities between the open ocean and sub-shelf water cavity are satisfied to a desired accuracy.
Similarly, the thin plate approximations (Eq.~\ref{eqn:UW_thinplate}) could be extended with additional terms to improve accuracy, for which (to our knowledge) there is no existing benchmark studies.
However, even for the shortest period considered in \S\,\ref{sec:results} ($T=10$\,s, i.e., $\omega\,/\,2\,\pi=0.1$\,Hz), the wavelengths in the shelf--cavity interval are much greater than the shelf thickness (Fig.~\ref{fig:wavenumbers}), indicating the thin plate approximations in Eq.~(\ref{eqn:UW_thinplate}) are likely to give accurate results.}

\new{
The dynamic problem ($\omega\neq{}0$) was considered, so that the derived approximation could be compared with the benchmark thin plate approximation, in order to identify the influence of water--ice coupling at the shelf front and extensional waves.
The static problem ($\omega=0$) can also be derived from the variational principle. 
In particular, the static problem couples the water and ice at the shelf front through the traction and bending moment conditions, where static water pressure provides the forcing.
However, the static problem requires a finite shelf, so that the resistive force at the grounding line (where the shelf meets land) prevents unbounded growth in the static displacements.}


The key finding that water--ice coupling and extensional waves have the greatest impact on ice shelf strains in the swell regime appears to contradict the findings of \citet{abrahams2023ocean} (see \S\,\ref{sec:intro}).
Similar to the numerical model of \citet{abrahams2023ocean}, the approximation derived in this study (\S\,\ref{sec:frequencydomain}) predicts extensional wave displacements that are greater than flexural wave displacements for low frequencies (long periods), and that the amplitude ratio becomes unbounded as frequency tends to zero (not shown)\new{, such that $\mathcal{T}^{\tn{(op$\rightarrow$ex)}} \to\infty$ as $\omega\to{}0$}.
This property is a consequence of the elliptical trajectories of water particles created by the incident waves having aspect ratios that increasing skew towards the horizontal axis as wavelengths increase.
\new{Our results show that extensional waves have a negligible impact on shelf strains for long periods (e.g., Fig.~\ref{fig:strain_profiles}).
Further, flexural waves hold greater energy than extensional waves for long periods, i.e., $\mathcal{T}^{(\tn{fl})}\gg\mathcal{T}^{(\tn{ex})}$ for $T\gg{}1$ (Fig.~\ref{fig:transmission}), where the small limiting values of $\mathcal{T}^{(\tn{fl})}$ are due to decreases in $\mathcal{T}^{\tn{(ex$\rightarrow$op)}}$ compensating for increases in $\mathcal{T}^{\tn{(op$\rightarrow$ex)}}$.
Intuitively, as incident waves get longer, the impact of the ice cover decreases, resulting in $\kappa\approx{}k$ (Fig.~\ref{fig:wavenumbers}) and most of the incident wave transmitting into a flexural-gravity wave in the shelf--cavity interval ($\mathcal{T}^{\tn{(op$\rightarrow$ex)}}\approx{}1$).}


The strain magnitudes presented in \S\,\ref{sec:results} are one to two orders of magnitude smaller for wave periods in the swell regime than in the infragravity wave regime.
However, swell amplitudes are typically much greater than infragravity wave amplitudes, such that the benchmark model predicts they create strains of comparable magnitude \cite{bennetts2022modeling}.
In particular, flexural-gravity waves at periods in the swell regime are amplified by crevasses in ice shelves \cite{bennetts2022modeling}, and, thus, our findings highlight a potential, additional role of water--ice coupling and extensional waves in these amplifications.

\section*{Acknowledgements}

LGB is supported by Australian Research Council grants FT190100404 and DP200102828.

\end{document}